\documentclass[a4paper,preprintnumbers,floatfix,twocolumn,aps,prb,unsortedaddress,superscriptaddress]{revtex4-2}

\usepackage[para]{threeparttable} 
\usepackage{amsmath}
\usepackage{graphicx} 
\usepackage{booktabs}
\usepackage{subcaption}
\usepackage{url}
\usepackage{hyperref}
\usepackage{miller}
\usepackage{bm}
\usepackage{enumitem}
\usepackage{color}
\usepackage[version=4]{mhchem}

\usepackage[paperwidth=210mm,paperheight=297mm,centering,hmargin=1.7cm,vmargin=2.5cm]{geometry}

\usepackage[utf8]{inputenc}

\begin{document}

\title{Nine-element machine-learned interatomic potentials for multiphase refractory alloys}

\author{Jesper Byggmästar}
\thanks{Corresponding author}
\email{jesper.byggmastar@helsinki.fi}
\affiliation{Department of Physics, University of Helsinki, Finland}
\author{Tiago Lopes}
\affiliation{Department of Physics, University of Helsinki, Finland}
\author{Zheyong Fan}
\affiliation{College of Physical Science and Technology, Bohai University, Jinzhou, China}
\author{Tapio Ala-Nissila}
\affiliation{MSP group, Department of Applied Physics, P.O. Box 15600, Aalto University, FIN-00076 Aalto, Espoo, Finland}
\affiliation{Interdisciplinary Centre for Mathematical Modelling and Department of Mathematical Sciences, Loughborough University, Loughborough, Leicestershire LE11 3TU, United Kingdom}

\date{\today}

\begin{abstract}
{New refractory alloys are being continuously designed and characterised for applications requiring good high-temperature mechanical properties and stability. Computational design from atomistic simulations is limited by interatomic potentials missing key elements, being too inaccurate, or computationally too slow for large-scale simulations. Here we present development of a refractory alloy database and two computationally efficient and general-purpose machine-learned potentials (tabGAP and NEP). We also design a cross-sampling strategy for effective sampling of training data using predictions from two potentials with completely different underlying architecture. The potentials support arbitrary alloy compositions of elements in groups four to six in the periodic table (Ti, Zr, Hf, V, Nb, Ta, Cr, Mo, W). The database is diverse yet multitargeted to enable simulations of refractory metals and alloys across different pure-metal, solid-solution, intermetallic, and glassy phases. We demonstrate the usefulness of the potentials by reproducing known pressure-, temperature-, and solute-induced phase transitions, grain boundary segregation, and simulations of radiation damage in the WTaCrVHf metallic glass.}
\end{abstract}

\maketitle

\section{Introduction}
\label{sec:intro}

The scope of alloy design has in recent decades expanded to include more and more elements in various compositions. This has been largely driven by the rapidly increased interest in novel concentrated multicomponent alloys, or medium- to high-entropy alloys, where several metals are combined in significant concentrations~\cite{tsai_high-entropy_2014,miracle_critical_2017-1,senkov_refractory_2010,miracle_highentropy_2025}. Compared to conventional alloys comprised of a base metal with minor concentrations of alloying elements, the design of concentrated alloys operates in a vast chemical space of possible alloy compositions. Navigating this chemical space in search of novel alloys presents challenges both experimentally and computationally. A key challenge in quantitative, atomistic-level modelling is the difficulty and complexity of the model development required to perform simulations. In particular, interatomic potentials that support simulations of a large number of elements in various combinations and concentrations must be developed, ideally with high enough accuracy to enable truly reliable simulations and predictions.

The emergence of machine-learned interatomic potentials (MLIPs) has, with little loss in accuracy, connected the scales of computationally heavy quantum-mechanics-based calculations with classical atomistic simulations that scale linearly with number of atoms~\cite{behler_generalized_2007,bartok_gaussian_2010,shapeev_moment_2016,drautz_atomic_2019}. Development of MLIPs for various materials of interest is now a common part of the toolbox in the large community of atomistic materials modelling. MLIPs for single elements or specific alloys and compounds can now be developed as a routine task with reasonable effort and tailored to the desired applications. In recent years widespread efforts have been put on development of universal MLIPs for materials across the periodic table, from hereon referred to as foundation potentials~\cite{chen_universal_2022,deng_chgnet_2023,batatia_foundation_2025}. A growing number foundation potentials are continuously developed with the prospect of enabling simulations of arbitrary materials. Nevertheless, their accuracy for novel materials or microstructural features (e.g., defects, surfaces, dislocations, phase transitions) that are not well covered by the training database remains poor or unknown. In light of this, we argue that there is still a large gap between targeted single- or few-element MLIPs and foundation potentials that deserves attention. While the poor accuracy of foundation potentials for targeted applications can be solved by fine-tuning approaches~\cite{radova_finetuning_2025}, they are still computationally too expensive for large simulations (millions of atoms) over relatively long time scales (nanoseconds)~\cite{batatia_foundation_2025}. A recent exception to this that aims to connect computational efficiency with transferability to arbitrary materials of the periodic table is NEP89~\cite{liang_nep89_2025}, although its accuracy and reliability for targeted simulations still requires validation and fine-tuning. On the other end, extending existing efficient and general-purpose single- or few-element MLIPs to include more elements requires significant effort.

Due to the important applications of high-entropy alloys, interatomic potentials that support many metallic elements are strongly sought-after yet still scarce whenever specific alloy compositions are of interest. Traditional interatomic potentials for many-element alloys include, among others, the embedded atom method (EAM) potentials by Zhou et al.~\cite{zhou_misfit-energy-increasing_2004}, the family of modified EAM (MEAM) potentials~\cite{choi_understanding_2018}, and the angular-dependent potentials by Starikov et al.~\cite{starikov_angulardependent_2026}. In the last five years, various MLIPs for many-element alloys have also emerged, such as several tabulated Gaussian approximation potentials (tabGAP)~\cite{byggmastar_modeling_2021,byggmastar_simple_2022,fellman_radiation_2025,byggmastar_segregation_2025}, moment tensor potentials (MTP)~\cite{yin_atomistic_2021}, spectral-analysis neighbor potentials (SNAP)~\cite{li_complex_2020}, neuroevolution potentials (NEP)~\cite{song_generalpurpose_2024,liu_utilizing_2025}, and atomic cluster expansion (ACE) potentials~\cite{wang_ductility_2025a}. Most of these potentials support only $4-6$ elements, which in practice limit them to specific high-entropy alloys with possible transferability to the corresponding lower-entropy alloys. 

The EAM potential by Zhou et al.~\cite{zhou_misfit-energy-increasing_2004} supports metals and alloys containing 16 different elements. Recently, a general-purpose MLIP that supports the same 16 elements using the NEP architecture was developed, named UNEP-v1~\cite{song_generalpurpose_2024}. Together, they offer flexible simulation of various alloys with high computational efficiency. Both the EAM and UNEP-v1 are, however, missing key elements of refractory alloys (Nb and Hf) and the applicability to simulations of various intermetallic and pressure- or temperature-stabilised phases is unclear. Crucially, Nb is part of both the original and still widely studied Senkov alloys MoNbTaVW and HfNbTaTiZr~\cite{senkov_refractory_2010,senkov_microstructure_2011}. Hf has also recently been shown to be a key element in tuning the properties of low-activation radiation-resistant alloys~\cite{el_atwani_quinary_2023}. Motivated by this and the need for general-purpose MLIPs to fill the gap to foundation potentials, we here develop a RHEA database and two MLIPs for the metals Ti, Zr, Hf, V, Nb, Ta, Cr, Mo, W and arbitrary alloys thereof. The RHEA database is a diverse and curated set of structures covering all relevant phases and compositions. We demonstrate the versatility of the MLIPs by simulating various alloys and phases in and far from equilibrium, including a million-atom metallic glass subjected to irradiation.

\section{Results}
\label{sec:results}

\subsection{RHEA training database}

\begin{figure*}
    \centering
    \includegraphics[width=0.7\linewidth]{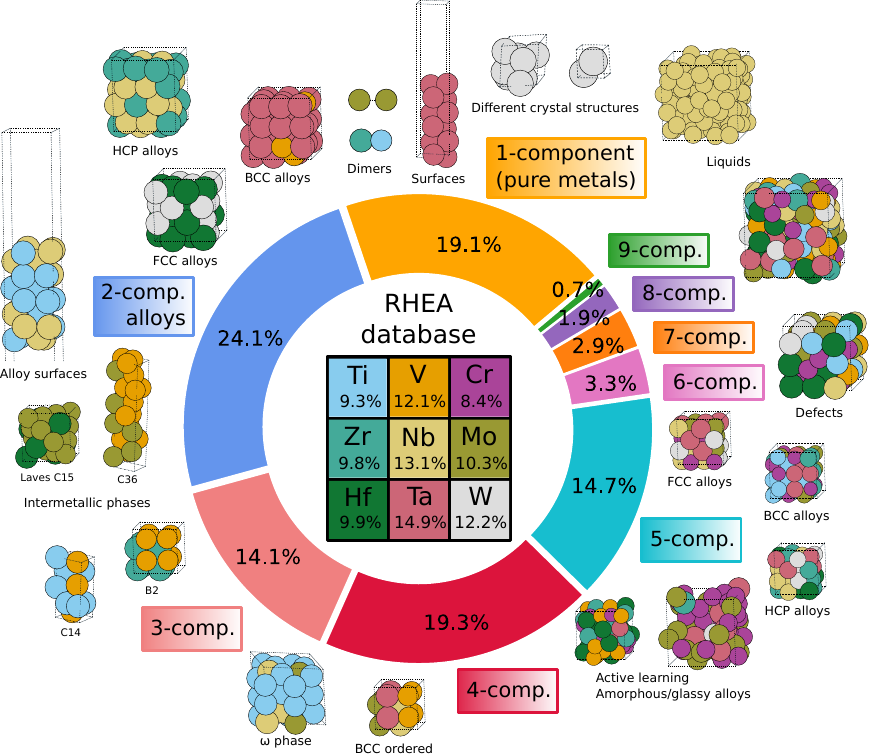}
    \caption{\textbf{RHEA training database}. Illustration of the content of the training database comprised of 22 477 structures and in total 694 078 atoms. The central element grid shows the atom fractions of each element and the outer circular chart shows the atom fractions of one- to nine-element structures along with representative snapshots of the various groups of structures.}
    \label{fig:db}
\end{figure*}

An illustration of the developed RHEA database is shown in Fig.~\ref{fig:db}. The starting point of the RHEA training database is the combination of all unique structures from the pure W, Mo, Ta, Nb, and V databases~\cite{byggmastar_gaussian_2020}, the Mo-Nb-Ta-V-W alloy database~\cite{byggmastar_modeling_2021,byggmastar_simple_2022}, the W-Ta-Cr-V database~\cite{byggmastar_segregation_2025}, and the Hf-Nb-Ta-Ti-Zr database~\cite{lopes_inpreparation}, all from our previous work. To populate the missing chemical and configuration space with data, we started by sampling binary alloys in the bcc phase at three different volumes and ten different compositions for all element pairs. A similar sampling was done for ternary alloys. In theory, sampling binaries and ternaries is enough to fill the feature space spanned by the local atomic environment descriptors used in tabGAP and NEP, since the former depends on at most three atoms and the NEP descriptors fundamentally depend on atom pairs~\cite{song_generalpurpose_2024}. For sampling the liquid phase we exploited this; however, for crystalline alloys we also included equiatomic compositions containing up to all nine elements for the three most relevant crystal structures (bcc, hcp, fcc) to ensure good accuracy and transferability to many-element alloys. 

The liquid phase is well sampled by the initial training databases for all metals and alloys included there. The missing chemical space was sampled by preparing molten alloys of all ternary equiatomic compositions at 0 GPa and 30 GPa using an intermediate version of the tabGAP (trained to all liquid data from the initial databases). Additionally we prepared molten nine-element alloys at ten different pressures from $-60$ to 100 GPa to ensure robustness of the MLIPs to extreme pressures and temperatures.

The RHEA database includes the known as well as hypothetical intermetallic phases of various compositions. These include the three Laves phases C15, C14, C36, the B2 phase, and the $\omega$ phase of pure and dilute Ti, Zr, and Hf alloys. We also enumerated all symmetrically distinct binary ordered phases in bcc lattices of $3-5$ atoms (1908 unique structures) using the icet package~\cite{angqvist_icet_2019}. All alloys from the Materials Project database were also included in the RHEA database (except a few artificial 1D structures)~\cite{jain_commentary_2013}. Additional coverage of finite-temperature short- to long-range order is ensured by iterative training in which we equilibrate alloys in hybrid Monte Carlo molecular dynamics (MC/MD) simulations at different temperatures and volumes using intermediate versions of the MLIPs.

Since we aim for the RHEA database to cover extreme conditions in terms of temperature, pressure, but also irradiation, it is essential to sample and ensure accurate and smooth interatomic repulsion at short distances. At very short distances, the screened Coulomb potentials that both MLIPs contain dominate (see Supplemental Information, SI), but the transitions to near-equilibrium interactions are not necessarily well-behaved and accurate unless guided by training data. For the pure metals this is done in the initial databases by including structures with unstable interstitial atoms in close (but not too close) contact with neighbouring atoms~\cite{byggmastar_machine-learning_2019}. To cover the missing repulsive interactions, we designed the following approach. In small bcc 16-atom random nine-element alloys, one atom is moved along the $\hkl<110>$ direction while keeping other atoms fixed to induce strong repulsion with neighbours. The $\hkl<110>$ direction is chosen because in the bcc lattice the atom approaches and passes between two neighbouring atoms, traversing an energy landscape with a saddle point of a few tens of eV repulsion~\cite{byggmastar_threshold_2024}. This corresponds to the energy range that defines the threshold displacement energy and more generally defect creation dynamics during irradiation, making it a crucial energy range to sample. We create training structures by enumerating and renaming the displaced atom (moved to the saddle point) and its two close-by neighbouring atoms to all element triplets of the nine elements (405 triplets). This creates a comprehensive set of structures that sample all pairs and triplets during the early stage of a irradiation-induced recoil event.

More details on the generation of the structures in the RHEA database are provided in the SI.

\subsection{Cross-sampling strategy for active learning}

In addition to the training-data generation methods described above we designed and employed a cross-sampling active learning strategy by comparing predictions of tabGAP and NEP. Active learning methods typically rely on uncertainty or extrapolation measures given by a single model if available, or uncertainties estimated from an ensemble of models. Here we find that a powerful strategy is to use the uncertainty given by the difference between predictions of two completely different MLIP architectures, in our case tabGAP and NEP.

The starting point for the cross-sampling is large pools of structures to sample from. We designed six pools of alloy structures that we deemed missing or poorly sampled in the manually created database: (1) elastic distortions, (2) point defects, (3) melt-quenched structures (metallic glasses), (4) melted structures, (5) MC/MD-optimised structures, and (6) surfaces. In all categories, we created a pool of a few hundred to a few thousand structures. Each structure is created as a random alloy composition containing a random number of elements (two to nine), with the desired modification (e.g., elastically and randomly distorting the cell, inserting vacancies or interstitial atoms), and then evolved in appropriate MD simulations (relaxation in $NVT$ or $NPT$ ensembles, melting and quenching, only melting, or MC/MD relaxation). In every pool of structures, half are simulated with tabGAP and half with NEP. Finally, the energies of all structures are computed with both MLIPs and sampling is done by picking the $N$ structures with largest differences in the MLIP total energy predictions. The cross-sampled structures are then computed in DFT and added to the training database. After retraining the MLIPs with the cross-sampled structures, the process was repeated in an iterative fashion until the differences between the predictions dropped to acceptable magnitudes (requiring four iterations in our case). In the SI, Fig. S1 shows how the predictions between the MLIPs drop with iteration.

During the cross-sampling we monitored how the true DFT energy of each structure deviated from the two MLIP energies. We found that, in general, the DFT energy lay between the two MLIPs. Typically, the MLIP that was used to simulate structure underestimates the DFT energy, and a cross-sampled structure is one where the other MLIP matches or clearly overestimates the energy. This highlights the power of the cross-sampling strategy: it eliminates both outlier structures corresponding to false or too deep local minima and outliers that are spuriously high in energy in either MLIP. The cross-sampling could also be combined with established model or ensemble uncertainty predictions but we leave that for future investigation.

\subsection{Machine-learned interatomic potentials}

\begin{figure*}
    \centering
    \includegraphics[width=\linewidth]{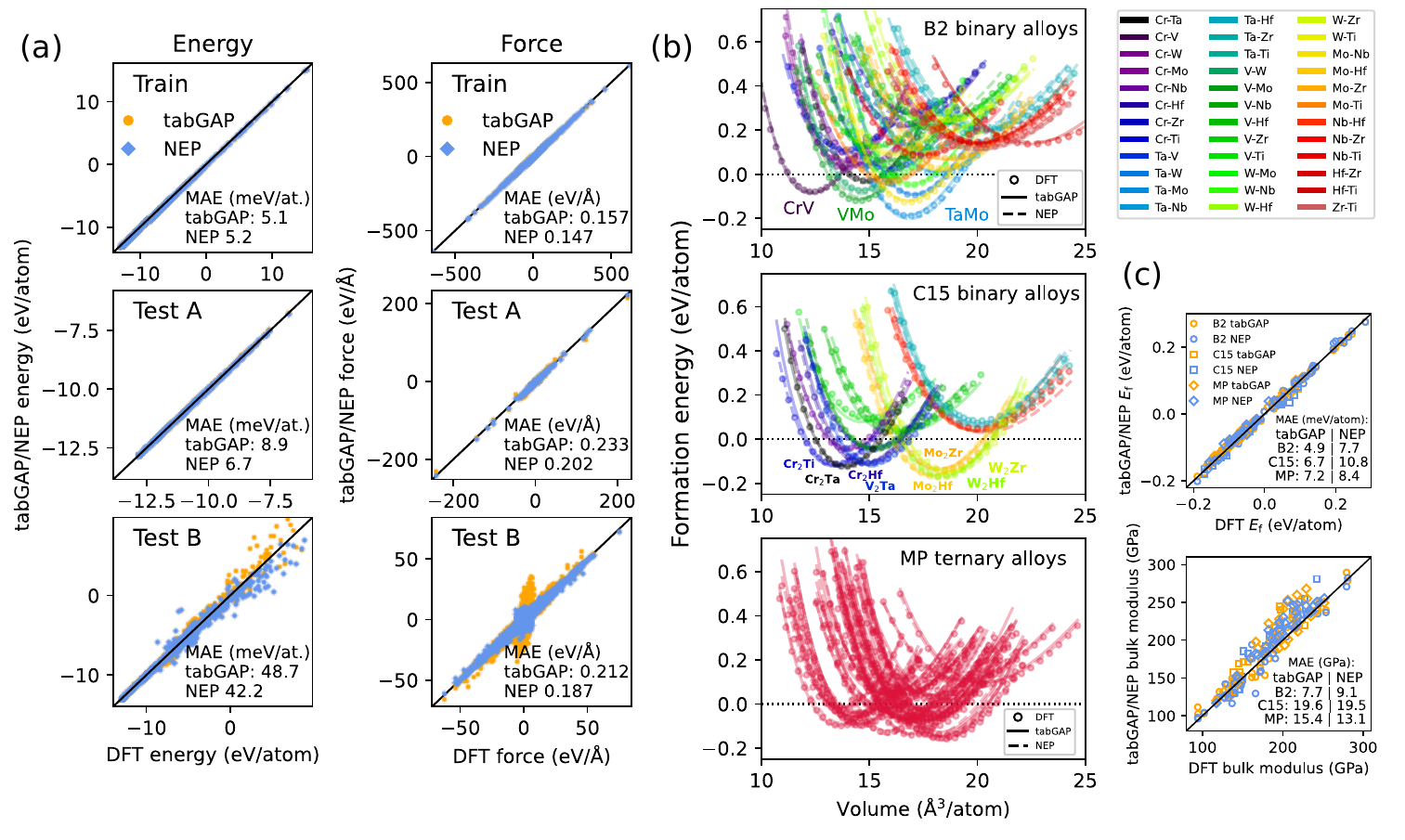}
    \caption{\textbf{Train and validation errors}. Column (a) shows parity plots and RMSEs for energies and force components for the training set, test set A (combination of the Mo-Nb-Ta-V-W test set from Ref.~\cite{byggmastar_simple_2022} with the W-Mo training data from Ref.~\cite{nikoulis_machine-learning_2021}), and test set B (subset of the training data from Ref.~\cite{song_generalpurpose_2024} that shares the same elements). Panel (b) shows energy-volume curves comparing the tabGAP and NEP to DFT for all binary B2 alloys, the most stable C15 Laves binary alloys, and all available ternary alloys from the Materials Project database. Panel (c) shows parity plots and RMSEs of the equilibrium formation energies and bulk moduli extracted from the energy-volume data in (b).}
    \label{fig:valid}
\end{figure*}

\begin{figure*}
    \centering
    \includegraphics[width=\linewidth]{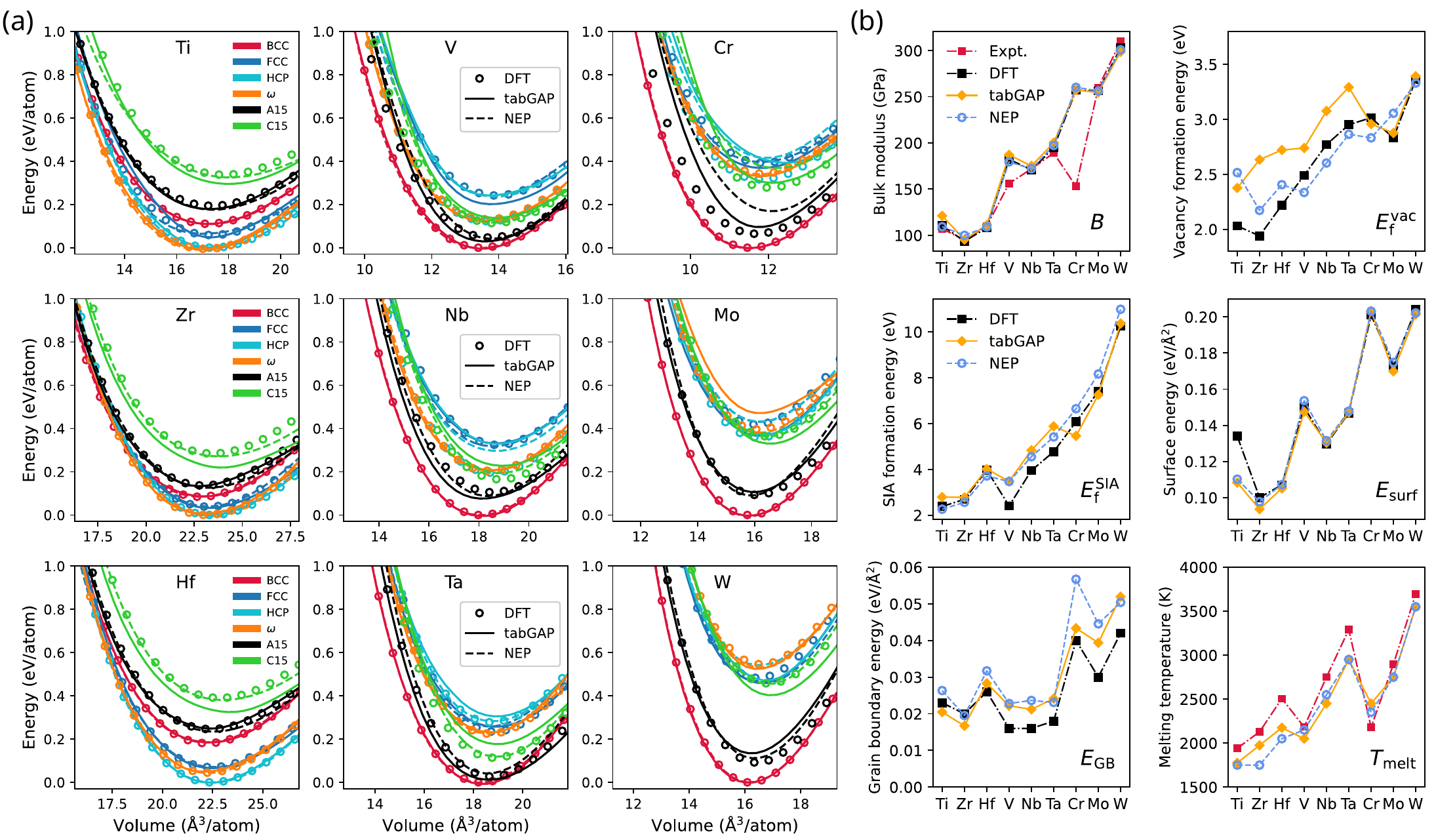}
    \caption{\textbf{Validation for the nine pure metals.} Energy-volume curves for isotropic volume-scaling of stable and hypothetical crystal phases, comparing tabGAP and NEP to DFT. The right column shows various material properties compared to DFT and experimental data.}
    \label{fig:pure}
\end{figure*}

The tabGAP and NEP MLIPs were iteratively improved during the construction of the RHEA database. Train and test errors of the final versions are shown in Fig.~\ref{fig:valid}(a), revealing similar accuracy for both MLIPs. The computational speed of both MLIPs is similar or faster than other MLIP frameworks, namely one order of magnitude slower than EAM and comparable to modified EAM (see SI for benchmark results). The train and test set A errors are low for both MLIPs, especially considering the wide ranges of energies and forces. Test set A is combination of the Mo-Nb-Ta-V-W test set from~\cite{byggmastar_simple_2022} with the W-Mo training data from~\cite{nikoulis_machine-learning_2021}, containing mainly bcc alloys (with and without defects) and liquids. Interestingly, the NEP performs better than tabGAP on the test B dataset, which are structures from the UNEP-v1 training data~\cite{song_generalpurpose_2024}. This suggests that either NEP performs better in extrapolation to structures far from the training data, or that NEP is more accurate because the structures were initially created by simulations with another NEP (UNEP-v1~\cite{song_generalpurpose_2024}). The performance of tabGAP on the test B structures could naturally be improved by adding them or similar structures to the training database, but since the structures for which tabGAP (and NEP) predictions are poor are very high in energy and quite artificial, we decided not to include this step.

Figure~\ref{fig:valid}(b) shows validation tests for energy-volume relations for various alloys, comparing DFT to the tabGAP and NEP for the B2 phase of all binaries, the C15 Laves phase for the most stable binary compositions, and all ternary alloys from the Materials Project database. From the minima and curvature of the energy-volume data, we compute the formation energies and bulk moduli and plot them against DFT in Fig.~\ref{fig:valid}(c). The results demonstrate several promising features of the MLIPs; the energy-volume predictions are smooth and well-behaved across wide volume and energy ranges, the order of the most thermodynamically stable phases are reproduced well in both MLIPs, and the formation energies and bulk moduli are fairly accurate in both MLIPs. Note that while B2, Laves, and Materials Project alloys are included in the training database, only one or a few near-equilibrium volumes are included.

Robust predictions for the pure metals provide a good basis for reproducing properties of arbitrary alloys, which is the main aim of the RHEA MLIPs. In Fig.~\ref{fig:pure}(a) we benchmark properties of all nine pure metals predicted by the MLIPs and compared against DFT and, when possible, experimental data. Figure~\ref{fig:pure}(a) shows energy-volume curves of six crystal phases. Both MLIPs predict smooth and well-behaved equations of state with the correct relative stability for the crystal phases. The $\omega$ phase is correctly reproduced as the ground state at zero temperature in Ti and Zr. The largest discrepancies are seen for Cr. This is partly intentional. Since Cr has a complex magnetic ground state which is not considered in the training data or the MLIPs (due to lack of explicit magnetic degrees of freedom), we intentionally included less data for pure Cr. For this reason the MLIPs are not physically accurate for pure Cr simulations, but for typical concentrated alloys containing Cr the magnetic moments are suppressed and the MLIPs are then reliable~\cite{zhao_defect_2020,byggmastar_segregation_2025,el-atwani_helium_2020}.

Figure~\ref{fig:pure}(b) summarises various pure-metal properties: bulk moduli, point defect formation energies, surface energy and grain boundary energies of the most stable orientations, and melting temperatures. Overall, the MLIPs show good performance as compared to DFT and experiments. One outlier is again the bulk modulus of pure Cr, which is overestimated as compared to the experimental value corresponding to the true magnetic ground state of Cr. The lack of magnetism has a smaller effect on other Cr properties, such as the melting point and surface energy. Additional validation for pure-metal properties is provided in the SI.

\subsection{Phase diagrams and phase transitions}

\begin{figure}
    \centering
    \includegraphics[width=\linewidth]{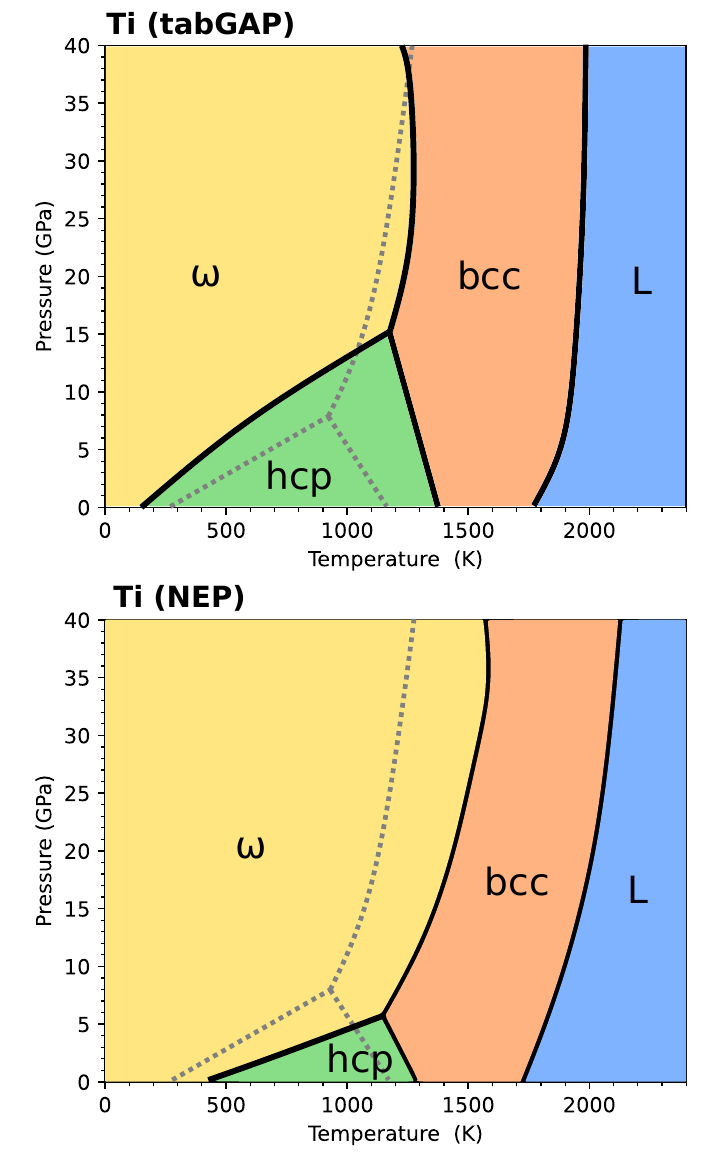}
    \caption{\textbf{Pressure-temperature phase diagrams of Ti}, drawn based on free-energy calculations and solid-liquid simulations using the tabGAP and NEP. The dotted lines show the phase boundaries drawn from experiments~\cite{dewaele_high_2015}.}
    \label{fig:PT}
\end{figure}

\begin{figure*}
    \centering
    \includegraphics[width=0.8\linewidth]{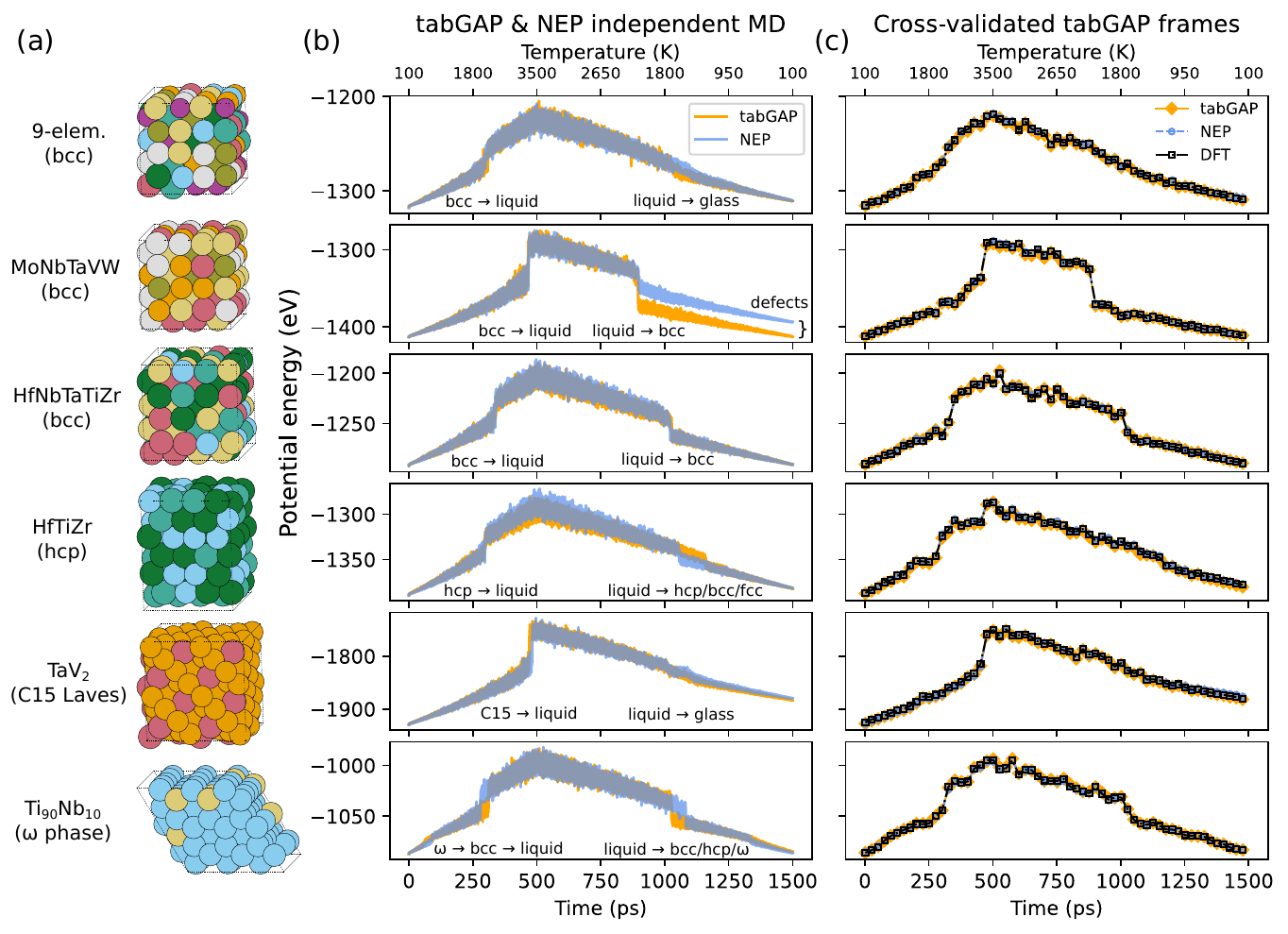}
    \caption{\textbf{Phase stability validation during heating and cooling.} Potential energy during heating and cooling of six different alloys to beyond the melting point and back to 100 K using the tabGAP and NEP. The various phase transitions occurring are indicated. Column (c) shows validation of both MLIPs as comparison of energies to DFT for structures sampled from the tabGAP trajectories.}
    \label{fig:melt-quench}
\end{figure*}

Figures~\ref{fig:valid}-\ref{fig:pure} show that both MLIPs perform well in predictions of zero-temperature alloy stability and crystal phase energetics as well as various defect energies and melting points of the pure metals. Our next aim is to show that the RHEA database and the MLIPs can also reproduce relevant phase transformations as functions of temperature, pressure, and composition. We begin by showing the pressure-temperature phase diagram of pure Ti predicted by tabGAP and NEP in Fig.~\ref{fig:PT}. The phase diagram is mapped by performing free-energy calculations and two-phase coexistence simulations (see Methods). The phase diagrams of pure Zr and Hf are also mapped in the same way and shown in the SI.

Figure~\ref{fig:PT} shows that both MLIPs predict qualitatively correct phase diagrams when compared to the approximate experimental phase transition lines from Ref.~\cite{dewaele_high_2015} drawn in grey dashed lines. The tabGAP somewhat overestimates the triple point and the size of the hcp stability region, while NEP slightly underestimates them and shows a systematic offset in the $\omega$-hcp and $\omega$-bcc transition lines compared to the experimental lines. Both MLIPs overestimate the hcp-bcc transition temperatures by $100-150$ K at low pressures. Both MLIPs predict increasing melting points with pressure, as is typical for metals. Experimental data for the melting line are available but vary so extremely that it is not possible to use for validating either MLIP~\cite{stutzmann_highpressure_2015}. For Zr and Hf (see SI), the phase diagrams are also qualitatively similar except for that NEP predicts the fcc phase to be the most stable Hf phase at extreme pressures ($>50$ GPa). For targeted applications, the accuracy of the phase diagram could likely be fine-tuned by sampling more training data at the pressures and temperatures of interest. Nevertheless, the results here show that the RHEA database already leads to reasonable phase diagrams, something that requires careful fitting to achieve with traditional interatomic potentials~\cite{hennig_classical_2008,mendelev_development_2016}. More importantly, this provides an important foundation for reproducing phase stabilities as functions of chemical composition.

To investigate and validate the alloy phase stability predicted by the MLIPs, we design six different alloy phases and subject them to heating-cooling simulations in MD. The results are summarised in Fig.~\ref{fig:melt-quench}. The six alloys are: (1) the full nine-element system in the bcc phase, (2) the bcc MoNbTaVW Senkov alloy, (3) the bcc HfNbTaTiZr Senkov alloy, (4) the hcp HfTiZr alloy, (5) the C15 Laves phase TaV$_2$, and (6) the $\omega$ phase of Ti$_{90}$Nb$_{10}$. All alloys are solid solutions except for the TaV$_2$ Laves alloy. Starting from 100 K, each alloy is heated up to 3500 K during 500 ps at zero pressure in the $NPT$ ensemble, after which they are cooled back down to 100 K over 1 ns. All six alloys are simulated with both MLIPs. The alloys undergo (by design) several phase transitions during the heating-cooling simulations. The first three bcc alloys remain stable in the bcc phase until melting. During cooling, the nine-element is supercooled into a metallic glass in both MLIPs. Both Senkov alloys recrystallise into the bcc phase, either perfectly or with residual defects. The initially hcp HfTiZr alloy melts and recrystallises into a mixed hcp/bcc/fcc phase during cooling in both MLIPs. The TaV$_2$ Laves alloy melts and solidifies into a metallic glass during cooling. Finally, the initially $\omega$-phase Ti$_{90}$Nb$_{10}$ alloy undergoes several phase transitions in both MLIPs. First, it transforms to the bcc phase already at low temperatures, after which it remains stable until melting. During cooling, it recrystallises first into a mixture of bcc and hcp in both MLIPs and in NEP eventually back to $\omega$ (but not in tabGAP, presumably since $\omega$-Ti is more stable in NEP than in tabGAP, Fig.~\ref{fig:PT}).

The above simulations demonstrate the robustness and accuracy of the MLIPs in several ways. First, all observed phases and phase transitions are reasonable and expected. For example, the Senkov alloys are known as stable bcc alloys. That Ti$_{90}$Nb$_{10}$ transforms to bcc is known from the experimental phase diagram~\cite{murray_nbti_1981}. Additionally, melting results in a clear increase in potential energy (latent heat) in all cases and both MLIPs, and all final glassy phases are higher in energy than the initial crystalline phase like they should. Finally, to further validate the results, we sampled structures from the tabGAP MD trajectories and recomputed the energy using DFT and NEP for direct comparison. Figure~\ref{fig:melt-quench}(c) shows the energies as functions of time and temperature, where it is clear that the DFT energy is close to both the tabGAP and NEP energy for every sampled structure across all observed phases and temperatures.

\subsection{Zr-Nb phase stability}

\begin{figure*}
    \centering
    \includegraphics[width=0.8\linewidth]{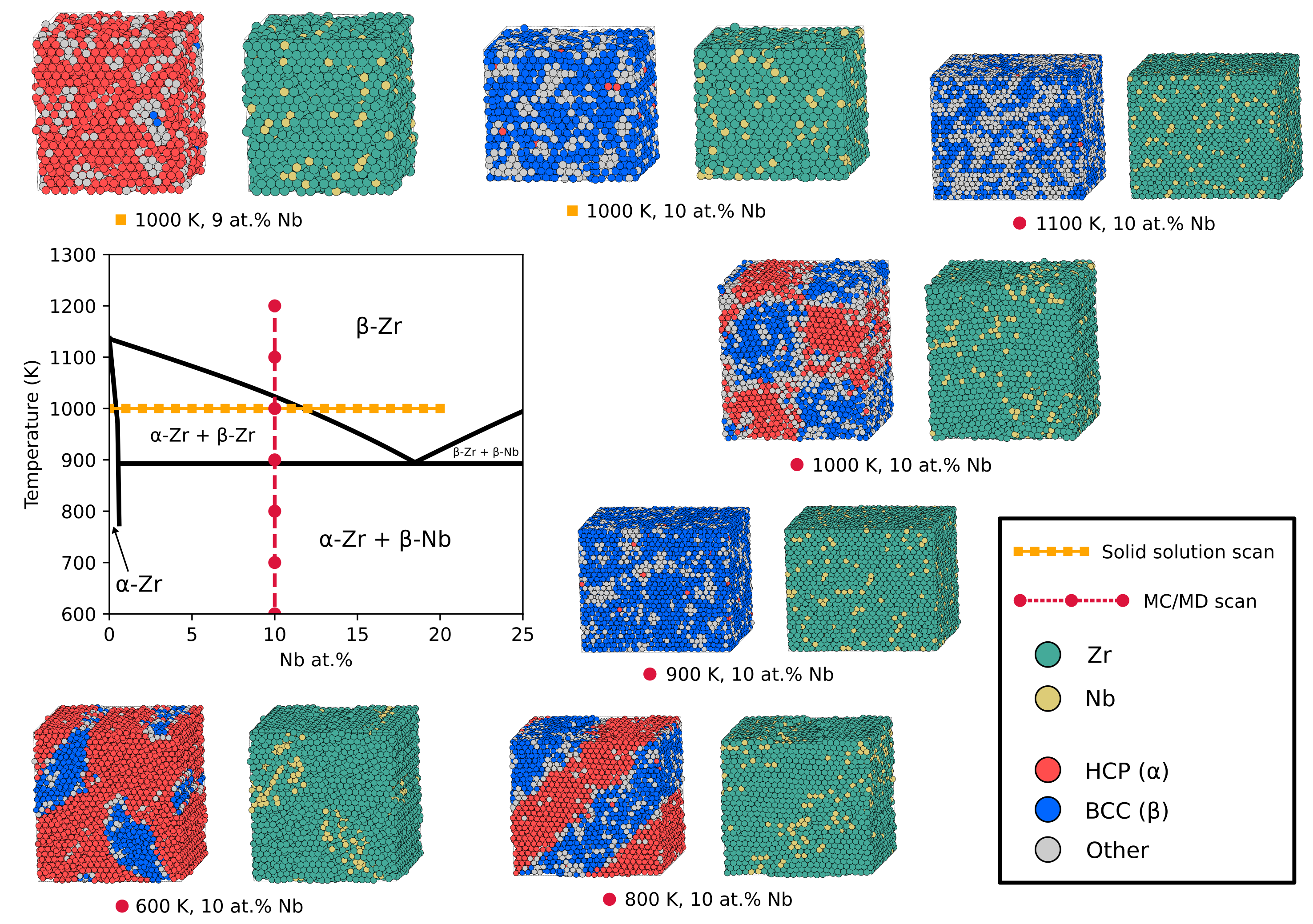}
    \caption{\textbf{Zr-Nb phase transitions.} Experimental phase diagram of Zr-Nb (redrawn from~\cite{abriata_nbzr_1982}) and snapshots of equilibrated systems from simulations at constant temperature (1000 K) and constant concentration (10 at.\% Nb). The constant temperature simulations are done with solid solutions in the $NPT$ ensemble (solid solution scan) with snapshots showing the transition from hcp Zr-10Nb to bcc Zr-10Nb solid solutions. The constant concentration simulations are done using the hybrid MC/MD method to produce the thermodynamically stable state (MC/MD scan). Snapshots from various temperatures are shown, illustrating the transition from two-phase (Zr-rich hcp and Nb-rich bcc) systems with increasing bcc stability to a single-phase bcc solid solution at high temperatures.}
    \label{fig:Zr-Nb}
\end{figure*}

As a case study of composition-dependent phase transitions predicted by the MLIPs, we focus on the binary Zr-Nb alloy system. Zirconium and titanium alloys often use bcc-stabilising elements like Mo or Nb in small concentrations to improve properties or promote the transition to the bcc ($\beta$) phase. Most notably, Zr-Nb alloys are used as cladding materials for nuclear fuel and a mixture of different stable and metastable phases have been observed depending on temperature, composition, irradiation, and synthesis method~\cite{tuli_thermodynamic_2023}. Figure~\ref{fig:Zr-Nb} shows the experimental Zr-rich part of the Zr-Nb phase diagram redrawn from Ref.~\cite{abriata_nbzr_1982}. We use the tabGAP to perform a number of MD equilibration simulations at constant temperature and constant composition to qualitatively investigate whether the composition- and temperature-induced phase transitions expected from the phase diagram can be reproduced. For constant temperature, we select 1000 K and start from hcp ($\alpha$) Zr-Nb solid solutions with increasing Nb content (orange line in Fig.~\ref{fig:Zr-Nb}). The experimental phase diagram shows that the $\alpha$-Zr solution should transition to the $\beta$-Zr phase as the Nb content increases, passing through a two-phase $\alpha$-Zr + $\beta$-Zr region. Our simulations predict that the critical Nb solid-solution concentration at 1000 K is 10 at.\%. At this and higher Nb content the lattice becomes dynamically unstable and transforms to the bcc phase, as shown by the snapshots in Fig.~\ref{fig:Zr-Nb}. This qualitatively agrees with the phase diagram that shows single-phase $\beta$-Zr to be the equilibrium phase at above 12 at.\% at 1000 K. 

For constant concentration, we select 10 at.\% Nb and run simulations at temperatures from 600 K to 1200 K (red lines in Fig.~\ref{fig:Zr-Nb}). These simulations start from an $\alpha$-Zr-10Nb solid solution and are done using the hybrid MC/MD method, which involves Monte Carlo swaps of atoms to produce possible thermodynamically favoured segregation and second-phase nucleation. At 10 at.\% Nb and lower temperatures, the experimental phase diagram shows that $\alpha$-Zr and $\beta$-Nb coexist. Indeed, at 600 K, the MC/MD simulation produces Nb segregation and nucleation of Nb-rich $\beta$-phase precipitates surrounded by almost pure elemental $\alpha$-Zr. As the temperature increases, the equilibrated $\beta$-phase regions are larger and include more Zr, until eventually the entire system becomes a single-phase $\alpha$-Zr solid-solution as illustrated by the snapshots in Fig.~\ref{fig:Zr-Nb}. This trend is consistent with the experimental phase diagram. Note that these MD simulations are limited in time and involve a degree of randomness that does not always lead to the same end result. For example, Fig.~\ref{fig:Zr-Nb} shows that at 800 K, the $\alpha$ and $\beta$ phases are equal in size, at 900 K it is fully $\beta$, while at 1000 K the simulations produce a two-phase mixture again. Only temperatures of 1100 K and above consistently produce complete $\alpha$-to-$\beta$ transitions, as they should based on the phase diagram.

\subsection{Grain boundary segregation compared to experiment}

\begin{figure*}
    \centering
    \includegraphics[width=0.8\linewidth]{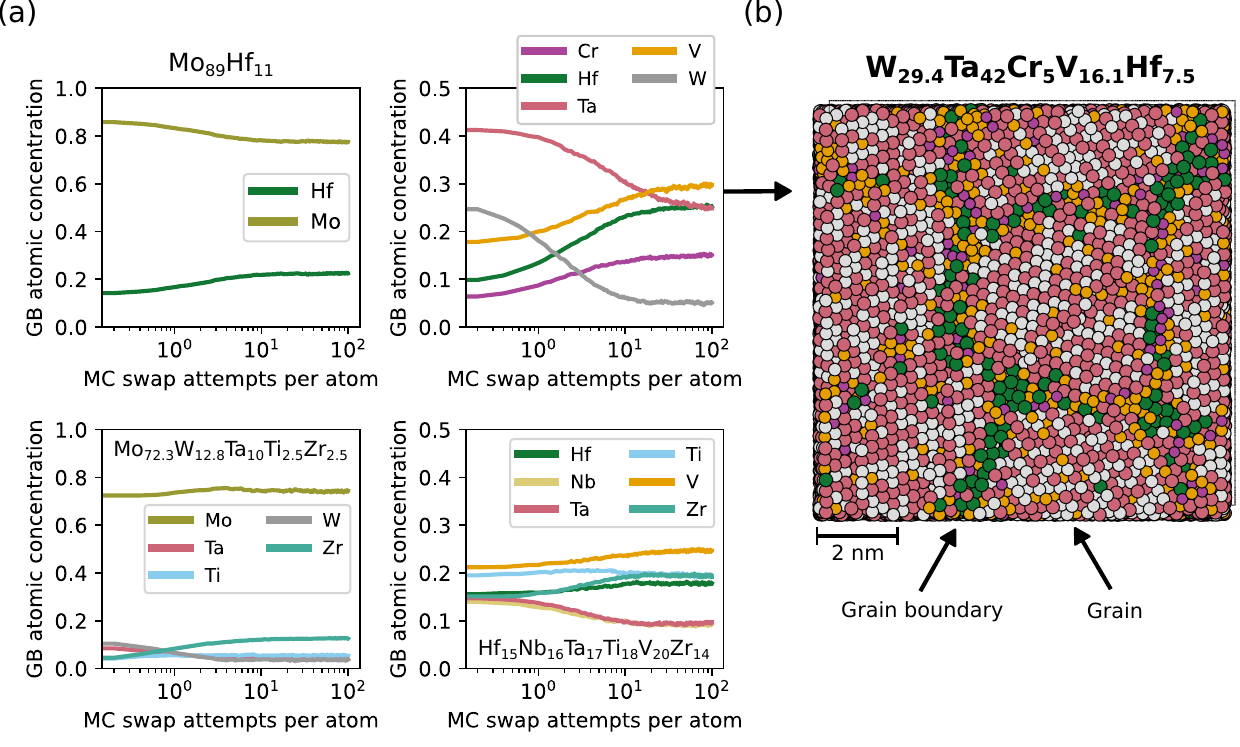}
    \caption{\textbf{Grain boundary segregation.} Panel (a) shows the change in atomic concentrations of grain boundary atoms in four experimentally studied alloys as functions of MC steps in MC/MD simulations with tabGAP at 1000 K. Panel (b) shows a snapshot of the final equilibrated \ce{W_{29.4}Ta42Cr5V_{16.1}Hf_{7.5}} alloy, with clear segregation to the grain boundaries.}
    \label{fig:GBseg}
\end{figure*}

In addition to simulations of phase stability and phase transitions, accurate and efficient MLIPs enable simulations of the role of microstructural defects. Synthesis of bulk samples of concentrated refractory alloys often results in segregation of certain elements to grain boundaries or a dendritic structure with compositional differences between dendrite and interdendrite regions. This can be driven by large differences in melting points between the elements, resulting in different recrystallisation temperatures, or thermodynamic driving forces to microstructural features such as grain boundaries~\cite{senkov_refractory_2010,li_segregationdriven_2019,ma_mechanism_2024,byggmastar_segregation_2025}. We collect four compositionally complex refractory alloys from the literature and compare the segregation trends observed experimentally with predictions from the MLIPs: \ce{Mo89Hf11}~\cite{senkov_exceptional_2025}, \ce{W_{29.4}Ta42Cr5V_{16.1}Hf_{7.5}}~\cite{el_atwani_quinary_2023}, \ce{Mo_{72.3}W_{12.8}Ta10Ti_{2.5}Zr_{2.5}}~\cite{ouyang_design_2023}, and \ce{Hf15Nb16Ta17Ti18V20Zr14}~\cite{gao_senary_2016}. All compositions are in atomic percent. For the simulations we use tabGAP and prepare polycrystalline systems with randomly oriented grains and grain boundaries for each alloy, which are then equilibrated in MC/MD simulations at 1000 K. This allows segregation to grain boundaries to naturally emerge, which can be qualitatively compared to the reported dendrite segregation in the experiments.

Figure~\ref{fig:GBseg} shows the concentration of elements in the grain boundary regions as functions of MC steps in the simulations. In all four cases, segregation occurs and the equilibrated grain boundary compositions differ noticeably from the initial global composition. In the binary \ce{Mo89Hf11} alloy, the simulation shows clear preferential Hf segregation to the grain boundaries ($22-23$ at.\% compared to the global 11 at.\%). In the experiment they reported Hf-rich channels with $16-24$ at.\% Hf, consistent with the simulated grain boundary concentration. In \ce{W_{29.4}Ta42Cr5V_{16.1}Hf_{7.5}}, the experiments also reported some Hf segregation. The simulation shows Hf segregation to be the strongest, but also predicts some V and Cr segregation to the grain boundaries.

In \ce{Mo_{72.3}W_{12.8}Ta10Ti_{2.5}Zr_{2.5}}, the experiments report preferential W and Ta segregation to dendrites, Mo and Zr to interdendrites, and Ti everywhere but a slight preference for interdendrites. The simulation also shows preferential Mo, Zr, and some Ti segregation to grain boundaries, leaving more W and Ta inside the grains. In \ce{Hf15Nb16Ta17Ti18V20Zr14}, the experiments report mainly Ta but also Nb to dendrite regions, with mainly Zr but also some Ti, V, and Hf to interdendrite regions. Similarly, the simulations predict Ta and Nb to remain in the grains, with mainly Zr but also some V and Hf to grain boundaries and Ti everywhere.

In all four cases, the simulations predict segregation trends that are consistent with those reported in experimental characterisations, even if the exact microstructures considered are different (nanocrystalline grains compared to dendritic structures). Simulating dendritic structures at realistic length scales is beyond the scope of the present work, but we believe the segregation trends are still comparable. Additionally, we also confirmed that the experimentally observed Nb and V segregation to grain boundaries in MoNbTaVW samples~\cite{liski_effect_2025} is reproduced with our MLIPs, as was also done previously~\cite{byggmastar_segregation_2025}.

\subsection{Radiation tolerance of WTaCrVHf metallic glass}

\begin{figure*}
    \centering
    \includegraphics[width=\linewidth]{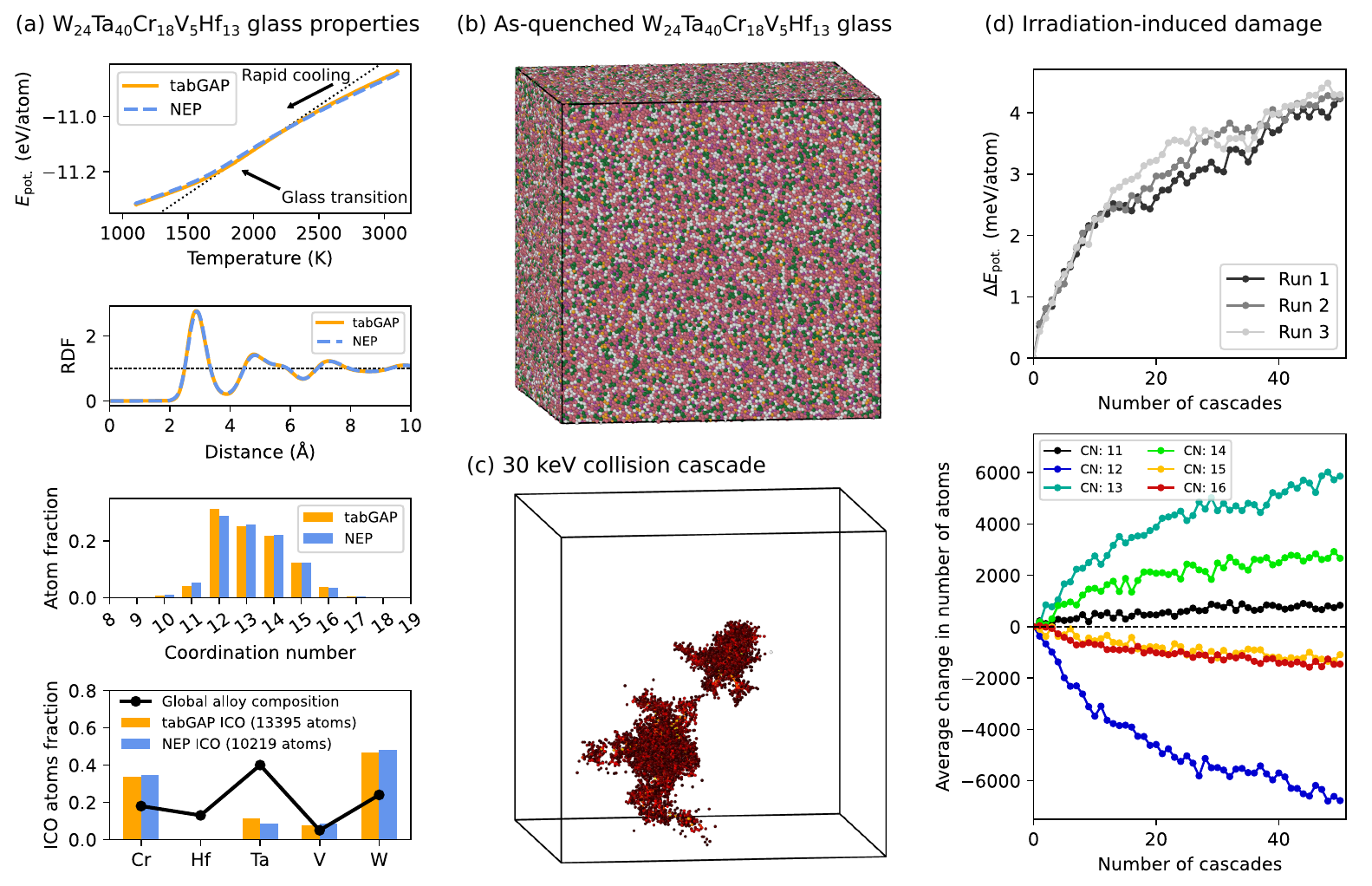}
    \caption{\textbf{One-million atom W$_{24}$Ta$_{40}$Cr$_{18}$V$_{5}$Hf$_{13}$ metallic glass}. Panel (a) shows the potential energy during cooling from the melt with formation of a glassy structure, the radial distribution functions (RDF) of the final glasses, the coordination numbers, and count and composition of icosahedrally coordinated atoms. All results are averaged over three independent simulations for each MLIP. Panel (b) shows a snapshot of a prepared glass at 1100 K. Panel (c) illustrates a 30 keV collision cascade by showing atoms with kinetic energies above 2 eV during the first 0.44 ps of a cascade simulation, with colours according to kinetic energy. Panel (d) shows the irradiation-induced damage as the change in average potential energy and average change in coordination numbers as functions of number of consecutive 30 keV cascades.}
    \label{fig:glass}
\end{figure*}

High-entropy metallic glasses are an emerging class of amorphous alloys. Tunes et al. prepared thin films of W$_{24}$Ta$_{40}$Cr$_{18}$V$_{5}$Hf$_{13}$ and found that they retained a stable glassy phase even during long-term annealing and high-dose irradiation at high temperatures (1173 K and 1073 K)~\cite{tunes_perspectives_2023}. To demonstrate the versatility of the MLIPs, we simulate the glass formation of W$_{24}$Ta$_{40}$Cr$_{18}$V$_{5}$Hf$_{13}$ from rapid cooling of the melt and subject the glassy systems to heavy irradiation exposure in consecutive collision cascade simulations. The results are summarised in Fig.~\ref{fig:glass}.

Figure~\ref{fig:glass}(a) shows various characterisation of the prepared glass system containing 1.024 million atoms. The potential energy as a function of temperature during rapid cooling at a rate 10$^{12}$ K/s from 3100 K to 1100 K shows a clear glass transition as a change in slope but no trace of recrystallisation. Both MLIPs produce very similar results both in the cooling behaviour and the local atomic structure of the glass. Figure~\ref{fig:glass}(a) also shows the radial distribution functions, the histogram of numbers of different coordination numbers of atoms (using a cutoff radius of 3.8 Å), and the number and composition of atoms in icosahedral environments as detected by the common neighbor analysis method in OVITO~\cite{stukowski_visualization_2010}. All results are averaged over three independent glasses in both MLIPs, obtained by starting from different random alloys. Atoms with coordination 12 are most frequent in the W$_{24}$Ta$_{40}$Cr$_{18}$V$_{5}$Hf$_{13}$ glass in both MLIPs. The composition of icosahedral atoms reveals that Cr and W are overrepresented as central atoms in icosahedra compared to the global alloy composition, while Hf- and Ta-centered icosahedra are rare and V-centered icosahedra similar to the global composition. These trends are consistent for the two MLIPs.

We subjected the prepared glassy alloys to heavy irradiation by simulating 50 consecutive 30 keV collision cascades with the tabGAP. Each collision cascade was initiated by giving a 30 keV recoil energy to a random atom close to and directed towards the centre of the simulation system. Figure~\ref{fig:glass}(c) shows an example of a collision cascade illustrated by all atoms with kinetic energies above 2 eV from the first $\approx 0.44$ ps of one simulation. The cascades in the metallic glass behave qualitatively similarly to high-energy cascades in crystalline metals and alloys, showing splitting into subcascades and heat spikes with extremely hot liquid-like cores~\cite{nordlund_primary_2018}. The cascade-induced melting and disorder rapidly recovers partially, leaving behind defects due to atoms being permanently displaced.

Figure~\ref{fig:glass}(d) quantifies the radiation-induced damage in the glass as the change in potential energy and change in different-coordinated atoms as functions of number of consecutive 30 keV cascades. No recrystallisation is observed during the irradiation, consistent with the experiments~\cite{tunes_perspectives_2023}, only subtle changes in the glass structure. The potential energy change is obtained by averaging over the last 6 ps of constant temperature and zero pressure relaxation of each simulation. The change in coordination numbers are averaged over the three independent simulation runs. The results in Fig.~\ref{fig:glass}(d) show that the heavy irradiation causes a small increase in average potential energy per atom, amounting to 4 meV/atom after 50 cascades. This translates to 4000 eV of energy stored in defects for the million-atom glass. The coordination analysis reveals that the radiation damage is mainly characterised by a decrease in the most favoured 12-coordinated atoms and a balancing increase in 13- and 14-coordinated atoms. The results in Fig.~\ref{fig:glass}(d) also indicate that the rate of radiation damage slows down with increasing irradiation dose. This is qualitatively similar to the saturation of damage widely observed both computationally and experimentally in crystalline metals and alloys~\cite{granberg_molecular_2021,wei_revealing_2024}. The main reason for this is that new cascades overlap with previously produced damage, causing less and eventually no new damage once the concentration of defects reaches a critical level (which depends strongly on the material and irradiation conditions~\cite{wei_revealing_2024,boleininger_microstructure_2023}).

\section{Discussion}

We have developed a diverse database of structures for refractory metals and alloys comprising the nine group $4-6$ elements of the periodic table and trained two computationally efficient MLIPs (tabGAP and NEP). The RHEA database was constructed aiming for a balance between diversity and targeted structures, so that the trained MLIPs are suitable for general-purpose use yet accurate for various specific intermetallic phases. The RHEA database is foreseen to become a robust piece of larger databases when, for example, fine-tuning foundation potentials for refractory alloys or when developing even larger many-element alloy MLIPs. The versatility and usefulness of the RHEA database and the trained MLIPs was demonstrated by reproducing basic materials properties and in particular various experimentally known phase transitions as functions of pressure, temperature, and alloy composition, including liquid and glassy phases. Both MLIPs also include accurate repulsive interactions at short distances through screened Coulomb potentials, enabling simulations of irradiation damage in arbitrary alloy compositions of the nine elements. As an example, we simulated radiation damage accumulation in the experimentally studied W$_{24}$Ta$_{40}$Cr$_{18}$V$_{5}$Hf$_{13}$ metallic glass, observing high stability of the glass phase as also seen in experiments.

Our work also shows how training and using two different MLIPs based on completely different architectures is a powerful strategy during all stages of MLIP development and use, from training data generation to validation and assessing the reliability of simulation results. For generating training data, we extensively used a cross-sampling strategy by sampling structures for which the two different MLIPs disagree. This approach is a simple complement to other methods of MLIP uncertainty quantification and active learning, which typically rely on ensembles or extrapolation measures within a single MLIP framework. When applied in simulations, the validity of the results can be reassured by cross-checking that the two MLIPs produce similar results.

We believe that the developed MLIPs strike a practically useful balance between versatility, accuracy, and computational efficiency. This puts them in the wide and poorly explored middle ground between now routinely developed limited single- or few-element MLIPs and the growing number of large and extremely general-purpose foundation potentials. Both tabGAP and NEP are fast enough to enable cheap multimillion-atom simulations over nanosecond time scales with competitive accuracy. As such, the MLIPs provide valuable tools for simulations of refractory alloys within the rich chemistry of Ti--Zr--Hf--V--Nb--Ta--Cr--Mo--W compositions in near- and far-from-equilibrium conditions.

\clearpage
\section{Methods}

\subsection{Density functional theory calculations}

DFT calculations for training and testing data are performed using the \textsc{vasp} code~\cite{kresse_ab_1993,kresse_ab_1994,kresse_efficiency_1996,kresse_efficient_1996} with projector augmented wave potentials~\cite{blochl_projector_1994,kresse_ultrasoft_1999} (version 5.4 \texttt{\_pv} for Cr and Ta and \texttt{\_sv} for all other elements). All calculations use consistent input with the PBE GGA exchange-correlation functional~\cite{perdew_generalized_1996}, 500 eV plane-wave cutoff energy, 0.15 Å$^{-1}$ maximum $k$-point spacing on $\Gamma$-centred Monkhorst-Pack grids~\cite{monkhorst_special_1976}, and 0.1 eV Methfessel-Paxton smearing~\cite{methfessel_high-precision_1989}. The parameters are the same as in our previous work~\cite{byggmastar_modeling_2021}. No spin-polarisation is used, which means the correct magnetic ground state of pure Cr and dilute Cr alloys is not obtained. However, the relaxed energy differences between different magnetic orders of pure Cr and Cr alloys are small~\cite{byggmastar_segregation_2025}.

\subsection{Tabulated Gaussian approximation potential}

The tabGAP is, as established in our previous work~\cite{byggmastar_simple_2022,byggmastar_modeling_2021}, fitted by first training a GAP using the QUIP code~\cite{bartok_gaussian_2010} with only low-dimensional descriptors for the local atomic environments (two-body, three-body, and EAM density~\cite{byggmastar_simple_2022}) and prefitted screened Coulomb repulsive potentials as an external term (see SI). The low dimensionality allows the energy predictions to be computed by GAP and tabulated onto 1D and 3D grids and saved as precomputed tabGAP potential files. Evaluation is then done efficiently with cubic spline interpolations~\cite{byggmastar_modeling_2021,glielmo_efficient_2018,vandermause_--fly_2020} in an implementation for LAMMPS (\url{https://gitlab.com/jezper/tabgap}). 

The main GAP input used for generating the tabGAP are cutoff radii of 5 Å for two-body and EAM descriptors, and 4.1 Å for the three-body descriptor. The shorter three-body cutoff optimises computational speed without notably sacrificing accuracy. GAP uses sparse Gaussian process regression~\cite{bartok_gaussian_2010,bartok_gaussian_2015}. The number of sparse points were 20 for two-body and EAM terms. For the three-body descriptors, we used more sparse points for pure-element triplets (400) than for mixed-element triplets ($80-148$), since the training structures contain far more pure-element triplets than mixed. Note that while the number of sparse points directly affects the computational speed of a GAP, the speed of tabGAP is not affected since the GAP energies are precomputed and stored in files. Upper limits for the sparse points were in our case, however, practically limited by maximum size of arrays in the QUIP code. 


\subsection{Neuroevolution potential}

The NEP is trained using the GPUMD code~\cite{xu_gpumd_2025} with the same prefitted Coulomb repulsive potentials included during training as for the tabGAP (see SI)~\cite{liu_large-scale_2023}. NEP uses the separable natural evolution strategy to optimise the parameters of an artificial feed-forward neural network with a single hidden layer~\cite{fan_gpumd_2022}. The descriptors for the local atomic environment are separated into radial and angular (three-body, four-body, five-body) components and are mathematically similar but not identical to the atomic cluster expansion~\cite{drautz_atomic_2019}. During the iterative training data generation, we trained in parallel a large and a small NEP model, separated by the $n_\mathrm{max}$ parameters (4 4 for small, 8 8 for large). The $l_\mathrm{max}$ parameters are 4 2 1, the cutoff radius 5 Å and number of neurons 80 in both cases. During the final validation, we found that the large NEP model strikes a better balance between accuracy and speed (significantly more accurate than the small while not significantly slower). Hence, we here only presented the large NEP model and discarded the small.


\subsection{Molecular statics and dynamics simulations}

Molecular statics and dynamics simulations using the MLIPs are done with both LAMMPS~\cite{thompson_lammps_2022} (tabGAP and NEP) and GPUMD~\cite{xu_gpumd_2025} (NEP only). LAMMPS is used for all tabGAP simulations and small calculations with NEP. GPUMD is used for running the large-scale (metallic glass) simulations on GPU with the NEP.

\subsection{Free-energy calculations}

Free-energy calculations are done to map the pressure-temperature phase diagram of pure Ti, Zr, and Hf. We use the methods in detail described by Freitas et al.~\cite{freitas_nonequilibrium_2016}, specifically the Frenkel-Ladd (FL) method. The FL method allows computation of the Helmholtz free energy, with the contributions from vibrational entropy, through thermodynamic integration and an analytically known reference phase. For the solid phases, we use the Einstein crystal as reference following Freitas et al.~\cite{freitas_nonequilibrium_2016}. The required simulation parameters were decided from convergence tests. The switching time was 25 ps, the box size $6000-7000$ atoms (depending on phase), and time step 1 fs in all simulations. After obtaining the Helmholtz free energy $F$, the Gibbs free energy was calculated as $G = F + PV$.

The phase diagrams were mapped by brute-force sampling $(P, T)$ points with 2 GPa and 50 K increments and calculating the Gibbs free energy for every relevant crystalline phase. The systems were first equilibrated at the desired pressure and temperature in the $NPT$ ensemble, after which the free-energy calculations were done in the $NVT$ ensemble as required. The zero-temperature phase transitions were accurately determined by computing the enthalpy $H = U + PV$, where $U$ is the potential energy, after full box relaxations to different pressures.

The solid-liquid phase boundary was determined from conventional two-phase solid-liquid interface simulations~\cite{morris_melting_1994} at different pressures.

\subsection{Hybrid Monte Carlo molecular dynamics}

Hybrid MC/MD simulations for the Zr--Nb phases and grain boundary segregation were carried out using LAMMPS. The number of swaps was set to 1\% of atoms every 10 MD steps. The MD integration was done in the $NPT$ ensemble for $10^5$ time steps, which proved to be enough for the segregation or phase decomposition to complete. For the grain boundary segregation simulations, we started from cubic bcc random alloy systems of $10\times10\times10$ nm (67 000 atoms) with four randomly oriented grains of similar size, created with Atomsk~\cite{hirel_atomsk_2015}. Grain boundary atoms were identified using the grain segmentation algorithm in OVITO~\cite{stukowski_visualization_2010}.

\subsection{Radiation damage simulations}

The radiation damage in the metallic glass was achieved by simulating 50 consecutive collision cascades using LAMMPS. The primary knock-on energy was 30 keV in every simulation and was given to an atom 60 Å in a random direction from the centre of the box and directed towards the centre. This allows the collision cascade to evolve in the centre of the box, which is simulated in the $NVE$ ensemble, while a 8 Å thick region around the borders are in $NVT$ to dissipate heat. The cascade simulation used an adaptive time step to accurately follow the high-energy atom trajectories. Energy losses due to electronic stopping was not considered. The total simulation time for each cascade was 100 ps. After that, the entire system was relaxed for 10 ps in the $NPT$ ensemble to release stress and stabilise the temperature at the desired 1100 K. After this, the system was randomly shifted in all three dimensions, with atoms exceeding the periodic boundaries wrapped back into the simulation cell. This is done so that the next cascade is initiated in a new random position yet still in the centre of the cell.

\section*{Data availability}

The RHEA database of structures, input files for DFT and potential training, as well as the tabGAP and NEP potential files are openly available from Ref.~\cite{ZENODO}.

\section*{Acknowledgements}

J.B. and T.L. were supported by funding from the Research council of Finland through the OCRAMLIP project, grant number 354234. 
ZF was supported by the Science Foundation from Education Department of Liaoning Province (No. LJ232510167001).
T.A-N. has been supported in part by the Academy of Finland through European Union -- NextGenerationEU instrument grant no. 353298, and grants nos. 370057 and 373647.
Computational resources provided by CSC - IT Center for Science Ltd. are gratefully acknowledged.

%

\end{document}


\title{Supplemental information of: Nine-element machine-learned interatomic potentials for multiphase refractory alloys}

\author{Jesper Byggmästar}
\thanks{Corresponding author}
\email{jesper.byggmastar@helsinki.fi}
\affiliation{Department of Physics, University of Helsinki, Finland}
\author{Tiago Lopes}
\affiliation{Department of Physics, University of Helsinki, Finland}
\author{Zheyong Fan}
\affiliation{College of Physical Science and Technology, Bohai University, Jinzhou, China}
\author{Tapio Ala-Nissila}
\affiliation{MSP group, Department of Applied Physics, P.O. Box 15600, Aalto University, FIN-00076 Aalto, Espoo, Finland}
\affiliation{Interdisciplinary Centre for Mathematical Modelling and Department of Mathematical Sciences, Loughborough University, Loughborough, Leicestershire LE11 3TU, United Kingdom}

\date{\today}

\maketitle

\tableofcontents

\clearpage
\section{Training data description}

The main manuscript illustrates and briefly describes the content of the RHEA database. Here we provide additional details of the structure and generation of the data. The data are available as an extended \texttt{.xyz} file from the repository~\cite{ZENODO}, directly compatible as input for both (tab)GAP and NEP training. The structures are grouped into a number of \texttt{config\_type}s for convenience and for allowing to specify different weights on distinct groups of structures during training. Below, in raw text, is a summary of the \texttt{config\_type} keywords (given in the header lines of each structure), the number of structures of each \texttt{config\_type}, the (average) number of atoms per structure, the total number as well as per-element number of atoms in each \texttt{config\_type}. 

\begin{tiny}
\begin{verbatim}
config_type               N_boxes    avg N_ats_in_box     N_atoms_total   # per-element atom Counter
----------------------------------------------------------------------------------------------------
isolated_atom             9          1.00                 9               # {'Mo': 1, 'Nb': 1, 'Ta': 1, 'V': 1, 'W': 1, 'Cr': 1, 'Zr': 1, 'Ti': 1, 'Hf': 1})
dimer                     659        2.00                 1318            # {'Ti': 177, 'Zr': 167, 'Nb': 158, 'Hf': 151, 'Mo': 150, 'V': 146, 'Ta': 128, 'Cr': 123, 'W': 118})
pure_elem_lowE            4330       18.39                79622           # {'W': 13259, 'V': 13157, 'Mo': 13020, 'Nb': 13020, 'Ta': 12966, 'Cr': 4924, 'Hf': 3094, 'Ti': 3091, 'Zr': 3091})
pure_elem_highE           2777       9.21                 25583           # {'V': 4354, 'Ta': 4338, 'W': 4338, 'Mo': 4264, 'Nb': 4212, 'Ti': 1062, 'Zr': 1062, 'Hf': 1059, 'Cr': 894})
liquid                    599        114.04               68310           # {'Ta': 10663, 'W': 10328, 'V': 10201, 'Nb': 8906, 'Mo': 8611, 'Cr': 6727, 'Hf': 4443, 'Zr': 4246, 'Ti': 4185})
short_range               784        30.59                23985           # {'Ta': 4392, 'W': 4334, 'V': 4199, 'Mo': 2891, 'Nb': 2877, 'Cr': 2543, 'Zr': 934, 'Ti': 928, 'Hf': 887})
bcc_binary_alloys         1608       54.00                86832           # {'Hf': 13164, 'Ti': 12858, 'Zr': 12756, 'Nb': 11067, 'Ta': 11067, 'V': 6480, 'Mo': 6480, 'W': 6480, 'Cr': 6480})
bcc_alloys                2150       54.00                116100          # {'Ta': 20614, 'W': 16497, 'V': 16481, 'Nb': 15792, 'Mo': 11733, 'Cr': 9840, 'Ti': 8394, 'Hf': 8378, 'Zr': 8371})
defects                   430        114.51               49238           # {'Ta': 9258, 'W': 8904, 'V': 8393, 'Nb': 6268, 'Mo': 4901, 'Cr': 4323, 'Hf': 3247, 'Zr': 1974, 'Ti': 1970})
bcc_alloys_surface        204        57.45                11720           # {'V': 1877, 'W': 1791, 'Ta': 1534, 'Nb': 1387, 'Mo': 1362, 'Cr': 1321, 'Zr': 1021, 'Ti': 722, 'Hf': 705})
bcc_alloys_ordered        4750       12.05                57258           # {'W': 7418, 'Ta': 7321, 'V': 7192, 'Cr': 6825, 'Nb': 6076, 'Mo': 5966, 'Zr': 5488, 'Hf': 5488, 'Ti': 5484})
hcp_alloys                910        50.69                46128           # {'Ti': 7108, 'Ta': 7085, 'Nb': 7083, 'Hf': 7079, 'Zr': 7079, 'Mo': 2679, 'V': 2672, 'W': 2672, 'Cr': 2671})
hcp_binary_alloys         528        54.00                28512           # {'Hf': 6684, 'Ti': 6378, 'Zr': 6276, 'Nb': 4587, 'Ta': 4587})
fcc_alloys                518        32.00                16576           # {'Zr': 1900, 'Ta': 1896, 'Nb': 1889, 'Hf': 1887, 'Ti': 1877, 'V': 1789, 'Cr': 1781, 'Mo': 1779, 'W': 1778})
fcc_binary_alloys         180        32.00                5760            # {'Zr': 1320, 'Ti': 1300, 'Hf': 1240, 'Nb': 950, 'Ta': 950})
misc_alloys               1324       50.30                66592           # {'Zr': 10427, 'Hf': 9208, 'Cr': 9066, 'Ti': 7258, 'Mo': 6649, 'V': 6253, 'W': 6105, 'Ta': 6006, 'Nb': 5620})
intermetallics            570        16.93                9648            # {'Zr': 1625, 'Hf': 1620, 'Ti': 1549, 'Cr': 880, 'V': 876, 'Ta': 796, 'Nb': 795, 'W': 756, 'Mo': 751})
materials_project         147        6.03                 887             # {'Cr': 199, 'Zr': 125, 'Mo': 94, 'Ti': 90, 'Hf': 89, 'Nb': 84, 'V': 81, 'W': 77, 'Ta': 48})

All                       22477      30.88                694078          # {'Ta': 103650, 'Nb': 90772, 'W': 84856, 'V': 84152, 'Mo': 71331, 'Hf': 68424, 'Zr': 67863, 'Ti': 64432, 'Cr': 58598})
\end{verbatim}
\end{tiny}

Below are descriptions of the structures in each \texttt{config\_type}.

\begin{itemize}
    \item \texttt{isolated\_atom}: Isolated atoms in vacuum. Computed in VASP in spin-polarised calculations to get correct reference energies and thus correct DFT-predicted dimer bond energies and bulk cohesive energies.
    \item \texttt{dimer}: Dimers of all element pairs. Mainly includes repulsive parts down to a distance above which VASP is deemed reliable, which is assessed by comparison to all-electron DFT data~\cite{nordlund_repulsive_2025a}. Also includes the dimer minimum, but not the full dissociation curve as these DFT calculations are done without spin-polarisation while isolated atoms are spin-polarised which would lead to inconsistencies.
    \item \texttt{pure\_elem\_lowE}: Pure-element structures from previous work and some additional structures generated in this work. Low-energy structures, mainly the ground-state crystal structure with or without strain, thermal displacements, or point defects.
    \item \texttt{pure\_elem\_highE}: Pure-element structures from previous work and some additional structures generated in this work. High-energy structures, unstable crystal structures and extended defects (surfaces) or extremely strained ground-state crystals.
    \item \texttt{liquid}: Liquid structures at various densities and temperatures, both pure element and alloys of all compositions. Generated by iterative training of tabGAP (i.e., from MD with previous versions of the tabGAP) and from cross-sampling (see main text).
    \item \texttt{short\_range}: Small BCC crystals, both pure metals and alloys, with an unstable self-interstitial or Frenkel pair so that one or several atoms are very close to each other. For ensuring accurate repulsion at close distances and smooth transition to the external repulsive potentials (see section~\ref{sec:reppot}). Care is taken not to put atoms too close for VASP to become unreliable (1.0--1.5 Å depending on elements, 1 Å for small atom bonds like Cr-Cr and 1.5 Å for Hf-Hf).
    \item \texttt{bcc\_binary\_alloys}: Sampling of all binary alloys in the BCC structure, 10 compositions for each element pair (5, 10, 20, ..., 90, 95 at.\%), three volumes per composition (one according to Vegard's law and approximately $\pm$ 20 \% of that volume). Atoms are randomly displaced to generate unique local environments and nonzero forces, using a normal distribution with standard deviation 0.08 Å.
    \item \texttt{bcc\_alloys}: All BCC alloys beyond binaries. Most are created similarly to the binaries above. Some are relaxed in MD with previous versions of the tabGAP to generate realistic thermal displacements. The composition sampling covers all alloys from 3 to 9 elements, with some bias towards ternary alloys, both concentrated and dilute. Some structures are elastically distorted. Some MD-relaxed structures collapsed or transformed to another phase (e.g., if the BCC phase is not stable, which mainly occurs for compositions with majority HCP metals).
    \item \texttt{defects}: Various alloys and some pure metals containing defects (vacancies, self-interstitial atoms, or both). Most structures are generated by iterative training or cross-sampling, i.e., they are relaxed with some previous version of tabGAP or NEP.
    \item \texttt{bcc\_alloys\_surface}: Small set of BCC random alloy \hkl(100), \hkl(110), and \hkl(111) surfaces. Generated manually and by cross-sampling.
    \item \texttt{bcc\_alloys\_ordered}: Various ordered BCC alloys generated by self-made scripts and by the \texttt{icet} package as described in the main text.
    \item \texttt{hcp\_alloys}: Sampling of all equiatomic alloy compositions with lattice constants according to Vegard's law, otherwise generated in the same way as the BCC binary alloys. Also includes HCP alloys from previous work.
    \item \texttt{hcp\_binary\_alloys}: Binary HCP alloys from previous work~\cite{lopes_inpreparation}.
    \item \texttt{fcc\_alloys}: Similar to HCP above.
    \item \texttt{fcc\_binary\_alloys}: Similar to HCP binary above.
    \item \texttt{misc\_alloys}: Collection of alloys that do not fit well into any other category. Mainly alloys that collapsed or transformed to other phases (intentionally or not). Most structures are from the cross-sampling iterations, where many structures collapsed to glassy or mixed phases.
    \item \texttt{intermetallics}: Various intermetallic phases, both known and hypothetical. Laves C14, C15, and C36, A15 phase, and $\omega$ phases of Ti/Zr/Hf alloys.
    \item \texttt{materials\_project}: All alloys containing the included elements from the Materials Project database, excluding a few clearly artificial and high-energy structures.
\end{itemize}

\clearpage
\section{Cross-sampling convergence}

\begin{figure}[h]
    \centering
    \includegraphics[width=0.8\linewidth]{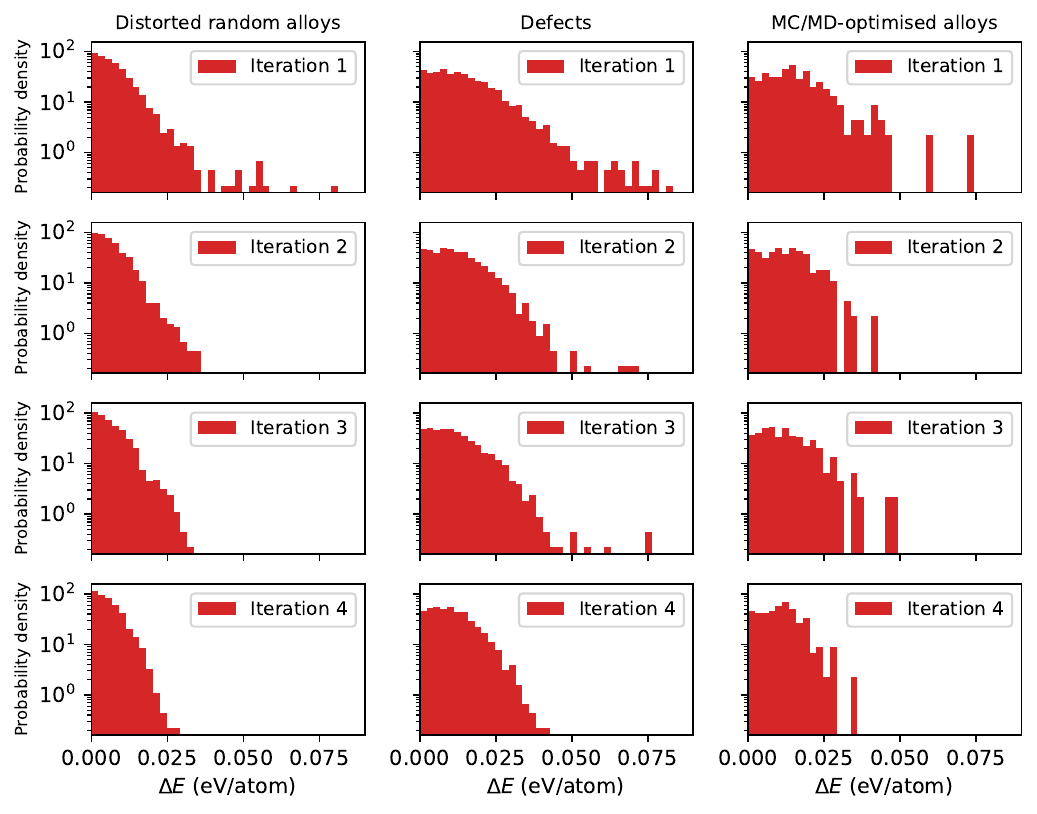}
    \caption{Distributions of the energy differences between tabGAP and NEP for three different pools of structures in the iterative cross-sampling strategy described in the main text.}
    \label{fig:SIcross}
\end{figure}

Figure~\ref{fig:SIcross} shows distributions of the energy differences between tabGAP and NEP for three different pools of structures used in the cross-sampling strategy described in the main text. The structures corresponding to the highest energy differences are picked, computed in DFT, and added to the training database, after which both MLIPs are retrained for the next iteration. The figure shows how the distributions become more narrow with each iteration with fewer and less extreme outliers (note the logarithmic scale). In the fourth iteration, we consider the MLIPs to agree to sufficient accuracy and pick the final sets of structures.

Example scripts implementing the cross-sampling are available in the repository~\cite{ZENODO}.

\clearpage
\section{Computational efficiency}

\begin{figure}[h]
    \centering
    \includegraphics[width=0.8\linewidth]{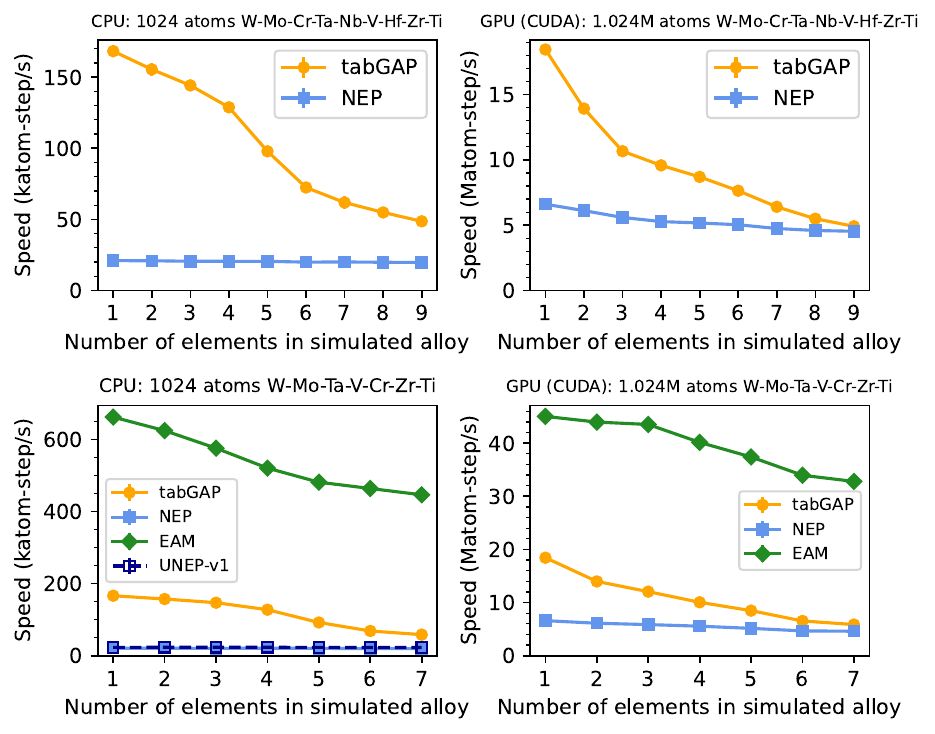}
    \caption{Computational CPU and GPU speed benchmark of tabGAP, NEP, UNEP-v1, and EAM (Zhou et al.) as a function of number of elements in random alloys. CPU simulations are done using LAMMPS and GPU using LAMMPS-KOKKOS for tabGAP/EAM and GPUMD for NEP. The results show the decrease in performance for tabGAP (and EAM) with increasing number of simulated elements in the alloy due to cache memory limitations and misses.}
    \label{fig:SIspeed_nelem}
\end{figure}

Figure~\ref{fig:SIspeed_nelem} shows the computational speed of tabGAP and NEP compared to EAM (Zhou et al.~\cite{zhou_misfit-energy-increasing_2004}) and UNEP-v1~\cite{song_generalpurpose_2024} in simulations of random alloys with increasing number of elements. The CPU benchmark simulations are done using LAMMPS on a single AMD Ryzen 7 PRO 5750G core. The GPU simulations are done using LAMMPS-KOKKOS for tabGAP and EAM and GPUMD for NEP on a single Nvidia Volta V100 GPU. All benchmarks are run for 1000 time steps in the $NVE$ ensemble with an initial temperature of 300 K. All tests were repeated three times to avoid and remove anomalous timings.

Figure~\ref{fig:SIspeed_nelem} shows that tabGAP is more efficient than NEP and UNEP-v1 but that the speed decreases with increasing elements. This is because tabGAP relies on storing and accessing large precomputed energy arrays during the 3D cubic spline interpolation. Since all energy arrays cannot be stored in the fast cache memory, increasing the number of elements (more energy arrays) increases the number of cache memory misses and slows down the simulation. This effect is minimised by a neighbour list sorting routine implemented previously in the LAMMPS-KOKKOS tabGAP implementation~\cite{byggmastar_four_2025} and in this work also in the CPU version, but the speed decrease is still significant. We also note that the decrease of the tabGAP efficiency with increasing elements depends on many factors (cache memory size, tabGAP grid sizes, alloy composition and order, cutoff radii, number of atoms in the simulation, etc.) and may differ from the results in Figure~\ref{fig:SIspeed_nelem} in other simulations and potentials. EAM potentials suffer from the same effect, but to a somewhat lesser degree due to smaller energy arrays. The speed of NEP is, on the other hand, almost independent of number of elements in the simulation.

\begin{figure}
    \centering
    \includegraphics[width=0.8\linewidth]{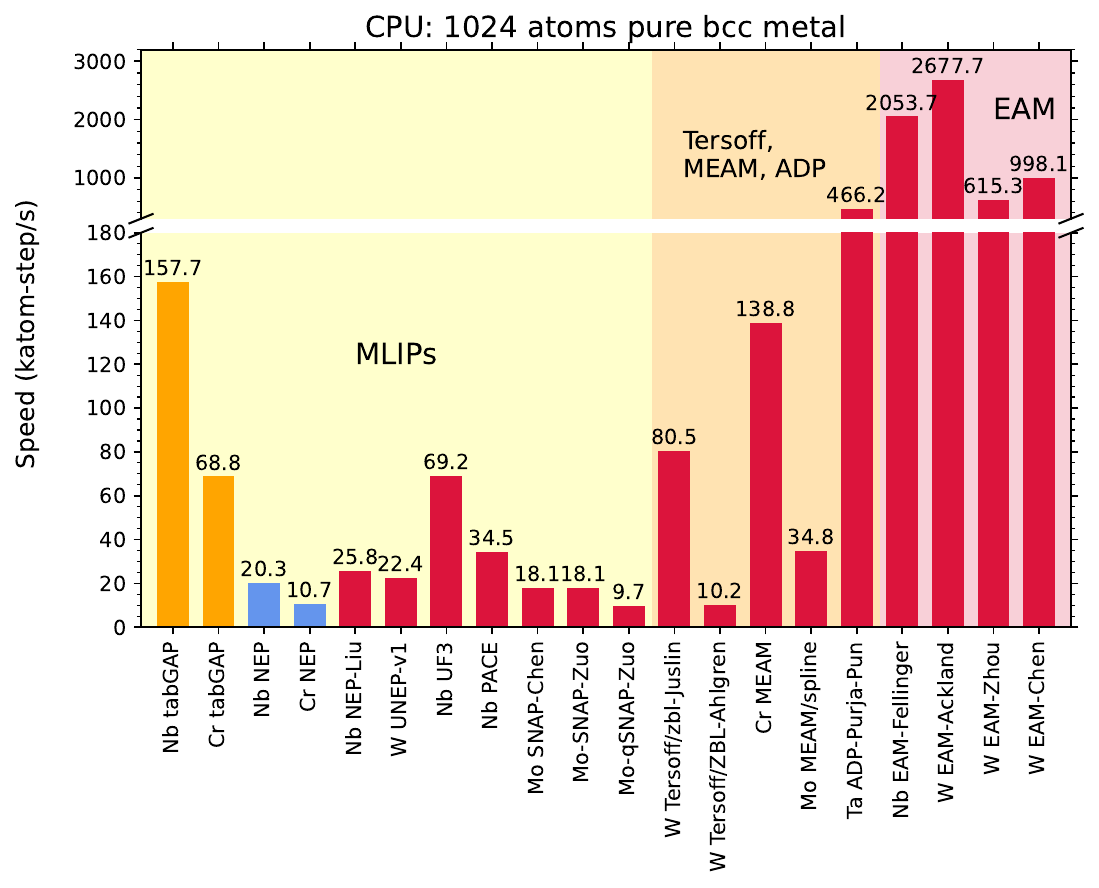}
    \includegraphics[width=0.8\linewidth]{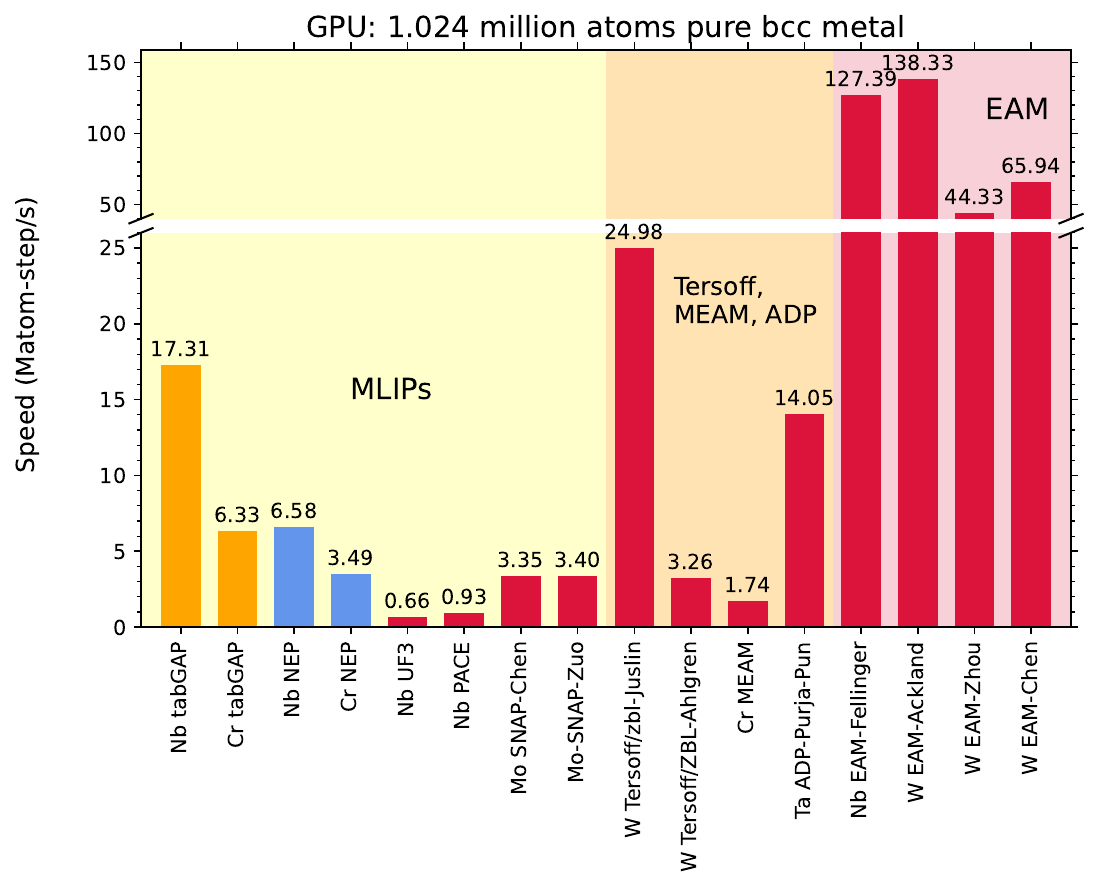}
    \caption{Computational speed benchmarks on CPU (top) and GPU (bottom) for a pure bcc metal with readily available interatomic potentials. The CPU simulations are done using LAMMPS on a single AMD Ryzen 7 PRO 5750G core. The GPU simulations are done on a single Nvidia Volta V100 GPU using LAMMPS-KOKKOS (if available), except for NEP which was run using the native GPUMD code. Note that the speed strongly depends on the number of neighbour atoms within the cutoff radii (which differ between potentials), which is illustrated by tabGAP and NEP results for both Nb (less atomically dense, i.e., fewer neighbours within cutoff radii) and Cr (atomically denser).}
    \label{fig:SIspeed_pure}
\end{figure}

Figure~\ref{fig:SIspeed_pure} shows speed benchmarks on CPU and GPU for tabGAP, NEP, and a number of other potentials readily available in a standard LAMMPS compilation. The potentials are grouped into three classes: MLIPs, traditional potentials with three-body dependence (Tersoff, modified EAM, angular dependent potential), and EAM. The tests in Figure~\ref{fig:SIspeed_pure} are all done on pure metals to allow comparison to many other potentials. Note, however, that the speed of a potential depends trivially on the number of neighbouring atoms within the cutoff radii. This is illustrated by showing tabGAP and NEP results for both Nb (less atomically dense, i.e., fewer neighbours within cutoff radii) and Cr (atomically denser). It is also clear from the fact that different EAM and Tersoff potentials show dramatically different speed.

In pure-element simulations, tabGAP is 6--8 times faster than NEP on CPU depending on number of neighbouring atoms within the cutoff radii. On GPU, tabGAP is only 2--3 times faster than NEP, highlighting that NEP is best used with its native GPUMD code. Compared to other potentials, tabGAP is significantly faster than all tested MLIPs (which only include the most efficient MLIPs) on both CPU and GPU. NEP is comparable to the other MLIPs on CPU but significantly faster on GPU.

Figure~\ref{fig:SIspeed_pure} also shows that the relative speed of potentials is different on CPU and GPU. For example, the ADP is very efficient and three times faster than tabGAP on CPU, but similar in speed on GPU. For Tersoff compared to tabGAP, the opposite trend is seen.

\clearpage
\section{Repulsive potentials}
\label{sec:reppot}

When training both the tabGAP and NEP, physically correct repulsion at very short interatomic distances is ensured by including external purely repulsive pair potentials as a baseline. The target energy for each training structure then becomes the difference between the DFT energy and this baseline repulsive energy. For the repulsive potentials not to interfere with near-equilibrium interatomic distances, the repulsive potentials are smoothly driven to zero by cutoff functions. This approach follows the methods described in more detail in Ref.~\cite{byggmastar_machine-learning_2019}.

The repulsive potentials are re-fitted Ziegler-Biersack-Littmark (ZBL) potentials~\cite{ziegler_stopping_1985} for each element pair:
\begin{equation}
    E_\mathrm{rep.} = \sum_{i<j}^N \frac{1}{4\pi \varepsilon_0} \frac{Z_i Z_j e^2}{r_{ij}} \phi_{ij} (r_{ij}/a) f_\mathrm{cut} (r_{ij}),
\end{equation}
where $a = 0.46848 / (Z_i^{0.23} + Z_j^{0.23})$ as in the universal ZBL potential. The screening function for an element pair AB is $\phi_\mathrm{AB}(x) = \sum_{i=1}^m \zeta_i e^{-\eta_i x}$. We used $m=3$ and fit the six parameters $\zeta_i$ and $\eta_i$ to the all-electron DFT data from Ref.~\cite{nordlund_repulsive_2025a} for each element pair. The cutoff function $f_\mathrm{cut} (r_{ij})$ is the same as in Ref.~\cite{byggmastar_machine-learning_2019} and smoothly forces the energy to zero in a pre-defined range.

The cutoff ranges and the fitted parameters are listed below as raw text. The columns are: cutoff start (Å), cutoff end (Å), $\zeta_1$, $\eta_1$, $\zeta_2$, $\eta_2$,  $\zeta_3$, $\eta_3$, $\zeta_4$, $\eta_4$. Each row is for one element pair. The order of rows (element pairs) is 1-1, 1-2, 1-3, ..., 1-9, 2-2, 2-3, ..., 8-8, 8-9, 9-9, where 1, 2, ..., 9 are the elements in alphabetical order: Cr Hf Mo Nb Ta Ti V W Zr. This format and the lines pasted below are exactly the format of the \texttt{zbl.in} input file for NEP training.

\begin{scriptsize}
\begin{verbatim}
1.0 2.0 0.62432709 0.52724872 0.17985857 1.59251784 0.19978689 2.32817531 0 0
1.0 2.2 0.46761483 0.88909152 0.21297607 2.98287067 0.31261612 0.40388501 0 0
0.8 2.2 0.20671801 9.44078152 0.49965657 1.26900977 0.428884 0.42626681 0 0
0.7 2.2 0.49811837 1.25665452 0.42783267 0.42607765 0.12694216 7.74046431 0 0
1.0 2.2 0.46700439 0.9213475 0.3270954 0.40805846 0.20402522 3.19206108 0 0
1.0 2.2 0.37440734 1.9039036 0.56883919 0.51978103 0.0507448 0.519783 0 0
1.0 2.0 0.1782988 2.29448513 0.19729121 1.64264918 0.62264504 0.52545345 0 0
1.0 2.2 0.32946711 0.40698282 0.19904867 3.3947085 0.47489525 0.93742154 0 0
1.0 2.2 0.50402738 1.22228461 0.10189284 6.41672406 0.41584423 0.42062546 0 0
1.0 2.2 0.59592609 0.56463409 0.3378634 2.39617239 0.06001117 0.25930197 0 0
1.0 2.2 0.17213842 3.69808253 0.31485428 0.37832137 0.51401042 0.97270952 0 0
1.0 2.2 0.51165313 0.90861154 0.29162012 0.36876844 0.18523058 3.23038014 0 0
1.0 2.2 0.5930524 0.56857818 0.33473184 2.39575589 0.06449729 0.26571147 0 0
1.0 2.2 0.25651787 0.37440487 0.21047215 2.9771018 0.5219775 0.84177329 0 0
1.0 2.2 0.21165315 2.979274 0.49082842 0.86742557 0.28862723 0.39247999 0 0
1.0 2.2 0.0704484 0.27305632 0.33301887 2.42079332 0.59005438 0.57393834 0 0
1.0 2.2 0.27621236 0.36224181 0.19530828 3.22049908 0.52075578 0.88265633 0 0
1.0 2.2 0.48796853 0.98194874 0.31579653 0.35741207 0.18798549 2.71874806 0 0
1.0 2.2 0.19020878 2.65056725 0.32076908 0.36142452 0.47789479 0.98036749 0 0
1.0 2.2 0.51344146 0.98119957 0.16477545 3.69293833 0.31816071 0.37914886 0 0
1.0 2.2 0.40970569 0.41443049 0.14048879 7.33390014 0.50897037 1.21591244 0 0
1.0 2.2 0.18396472 8.70291804 0.49897344 1.25387083 0.42609115 0.4247833 0 0
1.0 2.2 0.51684534 0.96288639 0.30904771 0.37485806 0.16779276 3.59037981 0 0
1.0 2.2 0.23908233 2.31575175 0.2905037 0.35143156 0.45403733 0.88232297 0 0
1.0 2.2 0.18049477 2.92908224 0.49502099 0.99695761 0.3227105 0.36116294 0 0
1.0 2.2 0.50494081 0.91672469 0.29923796 0.3728018 0.1826209 3.21613753 0 0
1.0 2.2 0.51944791 1.16631231 0.12504908 6.21792914 0.39038215 0.40495968 0 0
1.0 2.2 0.5026924 1.22895263 0.13951036 7.50424076 0.41849001 0.42091243 0 0
1.0 2.2 0.5085639 0.91518537 0.29670954 0.37129392 0.18666422 3.28455076 0 0
1.0 2.2 0.25827487 0.33608329 0.25427726 2.28636879 0.47267247 0.82744486 0 0
1.0 2.2 0.57289615 0.5763908 0.32394661 2.46358374 0.0739523 0.27722376 0 0
1.0 2.2 0.2028207 3.14399822 0.28344678 0.38591376 0.50676484 0.87847293 0 0
1.0 2.2 0.47408861 0.92079566 0.19860103 3.19599153 0.32249351 0.40575131 0 0
1.0 2.2 0.07833469 0.27911064 0.3294006 2.4745664 0.58841741 0.58199911 0 0
1.0 2.2 0.19519067 3.26708518 0.51520991 0.89414783 0.28453263 0.36642181 0 0
1.0 2.2 0.27696675 2.08564077 0.2225363 1.04949527 0.50214739 0.46974138 0 0
1.0 2.2 0.57377437 0.5002702 0.19290733 1.34666721 0.22937086 2.14239509 0 0
1.0 2.2 0.31065254 0.3962899 0.19280246 3.45766337 0.49736024 0.92468663 0 0
1.0 2.2 0.53564508 1.10903088 0.36371112 0.39206439 0.1189747 5.34000005 0 0
1.0 2.2 0.23582493 2.09349355 0.15563716 1.560913 0.60926898 0.51701385 0 0
1.0 2.2 0.47543791 0.94241702 0.19384435 3.38236532 0.33063109 0.40745887 0 0
1.0 2.2 0.10859677 6.0488931 0.51659195 1.17672301 0.39619262 0.41006409 0 0
1.0 2.2 0.32825 2.54931 0.09219 0.29182 0.5811 0.59231 0 0
1.0 2.2 0.29483836 0.37094151 0.19149279 3.38060865 0.51192616 0.91347455 0 0
1.0 2.2 0.25612795 2.34118951 0.49644544 0.8056594 0.23709107 0.32564545 0 0
\end{verbatim}
\end{scriptsize}

\begin{figure}
    \centering
    \includegraphics[width=\linewidth]{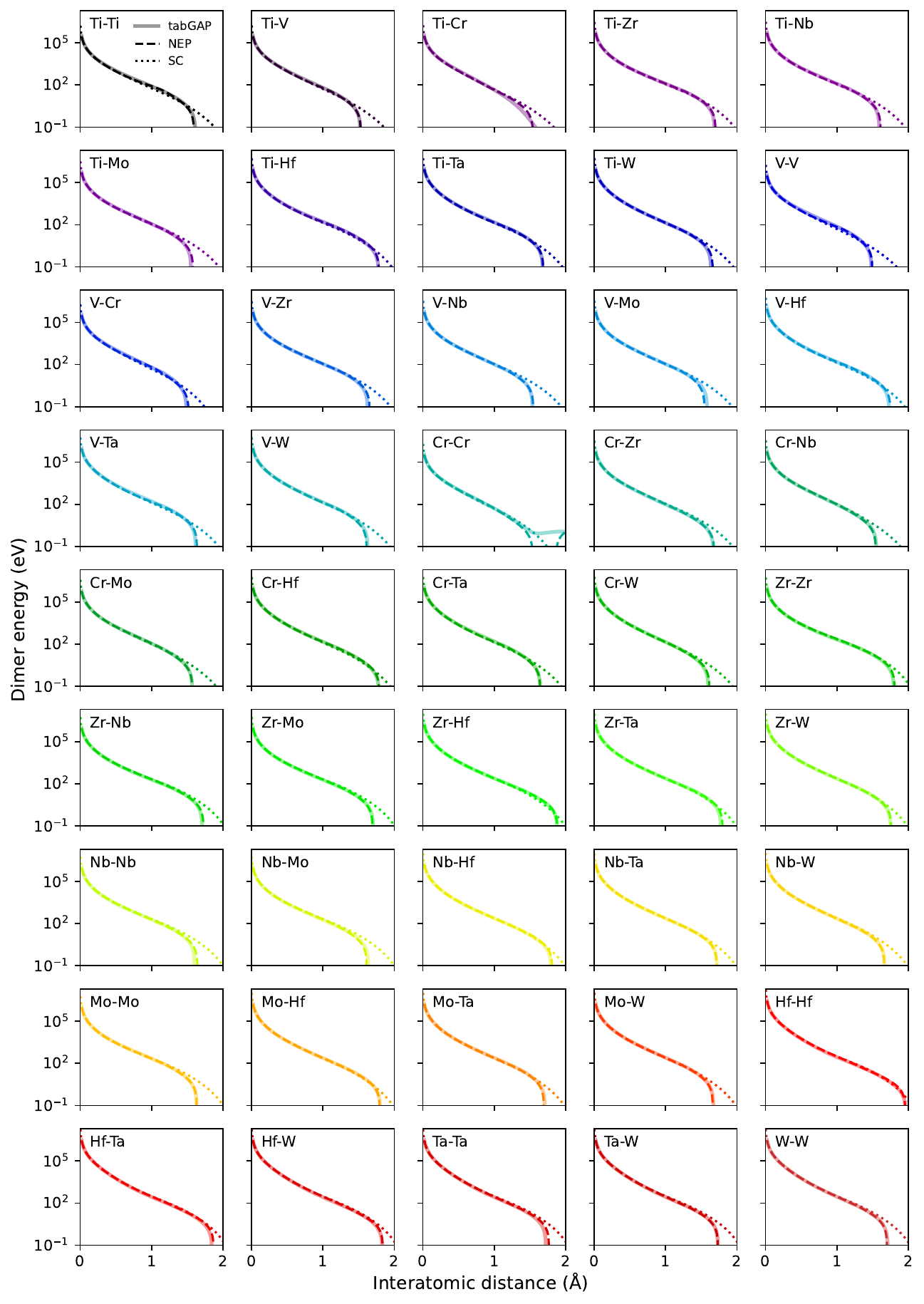}
    \caption{Repulsive parts of dimer curves for tabGAP, NEP, and the screened Coloumb potentials (SC) fitted to all-electron DFT data.}
    \label{fig:reppots}
\end{figure}

The fitted repulsive baseline potentials are plotted in Fig.~\ref{fig:reppots} (labeled ``SC'' as screened Coulomb). The same plot also shows the tabGAP and NEP energies, which are not identical to SC since there is also a nonzero machine-learned energy contribution. However, Fig.~\ref{fig:reppots} shows that this contribution is negligible at high energies as all potentials show similar energies. The figure also demonstrates that both tabGAP and NEP remain smooth and well-behaved even when interatomic distances approach zero.

\clearpage
\section{Pure metal properties}

\begin{figure}[h]
    \centering
    \includegraphics[width=0.8\linewidth]{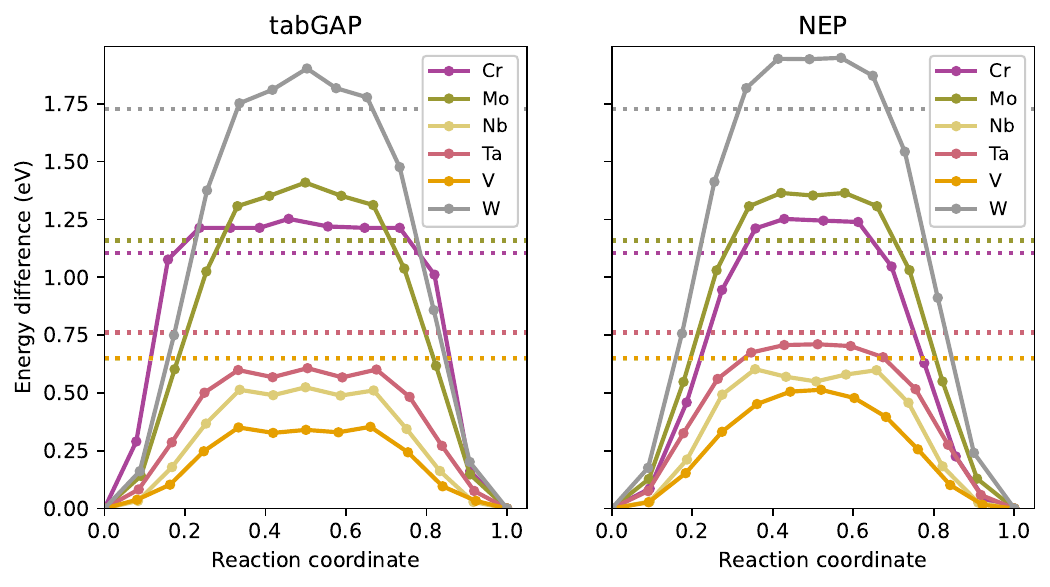}
    \caption{Vacancy migration energy barriers in the pure bcc metals computed with the NEB method. The dotted lines are the DFT migration energies from Ref.~\cite{ma_effect_2019}.}
    \label{fig:SInebbcc}
\end{figure}
\begin{figure}[h]
    \centering
    \includegraphics[width=0.8\linewidth]{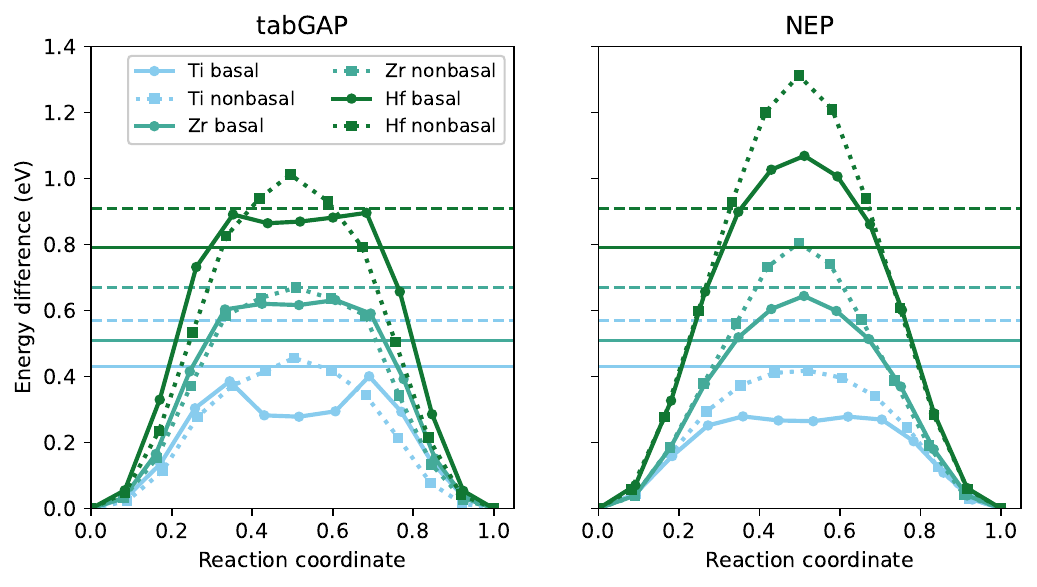}
    \caption{Vacancy migration energy barriers in the pure hcp metals computed with the NEB method for both migration in the basal plane (basal) and normal to the basal plane (nonbasal). The solid and dotted lines are the DFT migration energies from Ref.~\cite{verite_anisotropy_2007} for basal and nonbasal paths, respectively.}
    \label{fig:SInebhcp}
\end{figure}

Tables~\ref{tab:SI_pure}--\ref{tab:SI_vacmig} and Figures~\ref{fig:SInebbcc}--\ref{fig:SIPT} show various properties of the pure metals computed with tabGAP and NEP and compared to experimental and DFT data. Some observations regarding the performance of tabGAP and NEP are:

\begin{itemize}
    \item Elastic constants are overall reasonably accurate in both MLIPs, with some exceptions, and somewhat more accurate in NEP. Elastic constants of bcc metals are overall more accurate than for hcp metals. Elastic constants of Cr deviate from DFT and experiments mainly due to lack of spin-polarization in the training data. 
    \item Thermal expansion coefficients are either accurate or somewhat overestimated compared to experiment in both MLIPs. One noteworthy detail is that tabGAP captures the peculiar negative thermal expansion along the $c$ axis in Ti at low temperatures (below 300 K in tabGAP, below 170 K in experiments and DFT~\cite{souvatzis_anomalous_2007}), while NEP does not. The average linear thermal expansion, approximated as one third of the volumetric expansion, is still positive for Ti and all other metals.
    \item Vacancy migration energies are reasonably accurate in both MLIPs. The shape of the barriers sometimes differ and sometimes include a local minimum at the midpoint, which is also reported in DFT for the hcp metals~\cite{verite_anisotropy_2007}. Both MLIPs correctly predict the basal path to be favoured over the nonbasal path in the hcp metals.
\end{itemize}

\begin{table}[h]
    \centering
    \caption{Properties of the pure metals shown in Fig. 3 in the main text. The values for each property are given in the order DFT (or Expt. for $B$ and $T_\mathrm{melt}$)/tabGAP/NEP.}
    \begin{tabular}{lllllll}
    \toprule
    & $B$ (GPa) & $E_\mathrm{f}^\mathrm{vac}$ (eV) & $E_\mathrm{f}^\mathrm{SIA}$ (eV) & $E_\mathrm{surf}$ (eV/Å$^2$) & $E_\mathrm{GB}$ (meV/Å$^2$) & $T_\mathrm{melt}$ (K) \\
    \midrule
    Ti & 107/121/109 & 2.03/2.38/2.52 & 2.38/2.79/2.26 & 0.134/0.108/0.110 & 23/20/26 & 1941/1775/1750 \\
    Zr & 95/95/100 & 1.94/2.63/2.17 & 2.69/2.78/2.56 & 0.100/0.094/0.098 & 20/17/20 & 2128/1975/1750 \\
    Hf & 110/111/109 & 2.22/2.72/2.41 & 3.97/4.01/3.71 & 0.107/0.105/0.107 & 26/28/32 & 2506/2175/2050 \\
    V & 156/187/179 & 2.49/2.74/2.33 & 2.41/3.48/3.47 & 0.150/0.148/0.154 & 16/22/23 & 2183/2050/2150 \\
    Nb & 172/175/172 & 2.77/3.08/2.60 & 3.95/4.83/4.55 & 0.130/0.130/0.131 & 16/21/24 & 2750/2450/2550 \\
    Ta & 189/200/197 & 2.95/3.29/2.86 & 4.77/5.88/5.41 & 0.147/0.147/0.148 & 18/24/23 & 3290/2950/2950 \\
    Cr & 153/257/260 & 3.02/2.96/2.83 & 6.07/5.46/6.65 & 0.201/0.203/0.203 & 40/43/57 & 2180/2450/2350 \\
    Mo & 260/255/256 & 2.83/2.88/3.05 & 7.40/7.26/8.16 & 0.173/0.170/0.175 & 30/39/44 & 2896/2750/2750 \\
    W & 310/299/301 & 3.36/3.39/3.33 & 10.25/10.37/10.99 & 0.204/0.201/0.202 & 42/52/50 & 3695/3550/3550 \\
    \bottomrule
    \end{tabular}
    \label{tab:SI_pure}
\end{table}

\begin{table}
    \centering
    \caption{Elastic constants of the bcc metals at 0 K compared to DFT and room-temperature experimental data.}
    \begin{tabular}{lrrrr||lrrrr}
    \toprule
    & Expt.~\cite{rumble_crc_2025} & DFT~\cite{byggmastar_gaussian_2020} & tabGAP & NEP & & Expt.~\cite{rumble_crc_2025} & DFT~\cite{byggmastar_gaussian_2020,ma_universality_2019} & tabGAP & NEP \\
    \midrule
    V & & & & & Cr \\
    \midrule
    $C_{11}$     & 229 & 269 & 300 & 280 & $C_{11}$     & 340 & 448 & 444 & 527 \\
    $C_{12}$     & 119 & 146 & 157 & 130 & $C_{12}$     & 59 & 62 & 127 & 132 \\
    $C_{44}$     & 43 & 22 & 26 & 22 & $C_{44}$     & 99 & 102 & 83 & 113 \\
    \midrule
    Nb & & & & & Mo \\
    \midrule
    $C_{11}$     & 247 & 237 & 257 & 257 & $C_{11}$     & 464 & 468 & 462 & 475 \\
    $C_{12}$     & 135 & 138 & 135 & 133 & $C_{12}$     & 158 & 155 & 150 & 155 \\
    $C_{44}$     & 29 & 11 & 17 & 13 & $C_{44}$     & 109 & 100 & 97 & 110 \\
    \midrule
    Ta & & & & & W \\
    \midrule
    $C_{11}$     & 260 & 266 & 267 & 260 & $C_{11}$     & 522 & 521 & 519 & 516 \\
    $C_{12}$     & 154 & 161 & 170 & 161 & $C_{12}$     & 204 & 195 & 184 & 191 \\
    $C_{44}$     & 83 & 77 & 68 & 60 & $C_{44}$     & 161 & 147 & 138 & 139 \\
    \bottomrule
    \end{tabular}
    \label{tab:SI_Cxx_bcc}
\end{table}

\begin{table}
    \centering
    \caption{Elastic constants of the hcp metals at 0 K compared to DFT and room-temperature experimental data.}
    \begin{tabular}{lrrrr}
    \toprule
    & Expt.~\cite{rumble_crc_2025} & DFT~\cite{warwick_point_2025} & tabGAP & NEP \\
    \midrule
    Ti \\
    \midrule
    $C_{11}$     & 162 & 170 & 178 & 164 \\
    $C_{12}$     & 92  & 86 & 141 & 50 \\
    $C_{13}$     & 69  & 75 & 90 & 47 \\
    $C_{33}$     & 181 & 186 & 230 & 234 \\
    $C_{44}$     & 47  & 43 & 25 & 59 \\
    \midrule
    Zr \\
    \midrule
    $C_{11}$     & 143 & 141 & 136 & 141 \\
    $C_{12}$     & 73  & 68 & 78 & 82 \\
    $C_{13}$     & 65  & 66 & 71 & 75 \\
    $C_{33}$     & 165 & 164 & 192 & 177 \\
    $C_{44}$     & 32  & 26 & 26 & 32 \\
        \midrule
    Hf \\
    \midrule
    $C_{11}$     & 188 & 182 & 193 & 173 \\
    $C_{12}$     & 77  & 72 & 79 & 78 \\
    $C_{13}$     & 66  & 68 & 81 & 66 \\
    $C_{33}$     & 197 & 197 & 291 & 211 \\
    $C_{44}$     & 56  & 54 & 54 & 52 \\
    \bottomrule
    \end{tabular}
    \label{tab:SI_Cxx_hcp}
\end{table}

\begin{table}
    \centering
    \caption{Linear thermal expansion coefficients at room temperature in units of $10^{-6}$ K$^{-1}$ . Experimental values are from Ref.~\cite{rumble_crc_2025}.}
    \begin{tabular}{lrrr||lrrr||lrrr}
    \toprule
    & Expt. & tabGAP & NEP & & Expt. & tabGAP & NEP & & Expt. & tabGAP & NEP \\
    \midrule
     Ti & 8.6 & 11.7 & 11.5 & V & 8.4 & 8.4 & 11.6 & Cr & 4.9 & 8.3 & 7.6 \\
     Zr & 5.7 & 12.5 & 7.5 & Nb & 7.3 & 7.0 & 9.4 & Mo & 4.8 & 7.1 & 6.3 \\
     Hf & 5.9 & 8.8 & 7.1 & Ta & 6.3 & 5.6 & 7.4 & W & 4.5 & 5.6 & 5.5 \\
    \bottomrule
    \end{tabular}
    \label{tab:SI_thermal_exp}
\end{table}

\begin{table}
    \centering
    \caption{Monovacancy migration energies in eV. The two values for the hcp metals are for basal and nonbasal migration paths.}
    \begin{tabular}{lrrr||lrrr||lrrr}
    \toprule
    & DFT~\cite{verite_anisotropy_2007} & tabGAP & NEP & & DFT~\cite{ma_effect_2019} & tabGAP & NEP & & DFT~\cite{ma_effect_2019} & tabGAP & NEP \\
    \midrule
     Ti & 0.43/0.57 & 0.40/0.46 & 0.28/0.42 & V & 0.65 & 0.35 & 0.51 & Cr & 1.11 & 1.25 & 1.25 \\
     Zr & 0.51/0.67 & 0.63/0.67 & 0.64/0.80 & Nb & 0.65 & 0.52 & 0.60 & Mo & 1.21 & 1.41 & 1.36 \\
     Hf & 0.79/0.91 & 0.90/1.01 & 1.07/1.31 & Ta & 0.76 & 0.61 & 0.71 & W & 1.73 & 1.90 & 1.95 \\
    \bottomrule
    \end{tabular}
    \label{tab:SI_vacmig}
\end{table}

\clearpage
\section{Phase diagrams and free energies}

\begin{figure}[h]
    \centering
    \includegraphics[width=\linewidth]{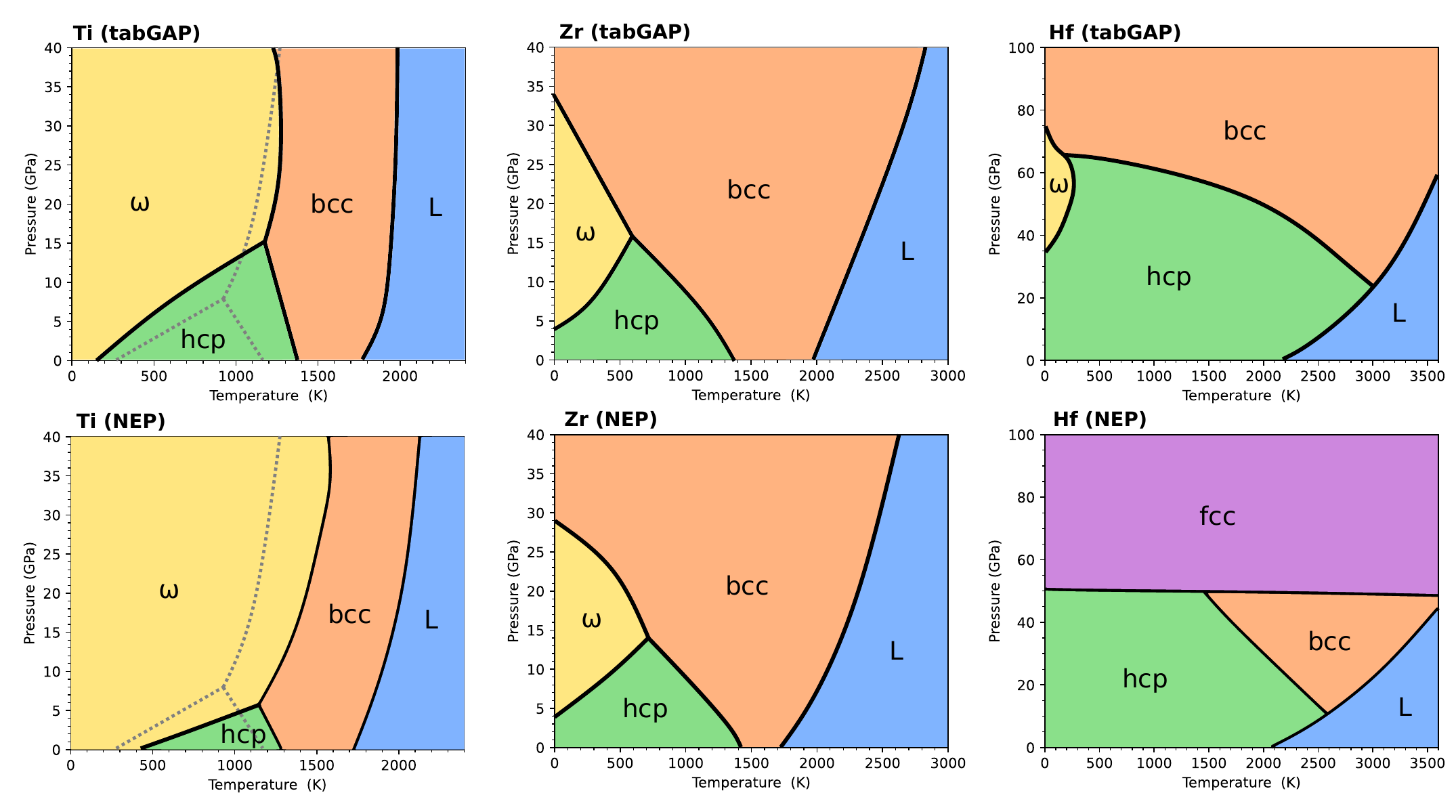}
    \caption{Phase diagrams for all three hcp metals predicted by tabGAP and NEP. See main text for methods.}
    \label{fig:SIPT}
\end{figure}

\begin{figure}
    \centering
    \includegraphics[width=0.8\linewidth]{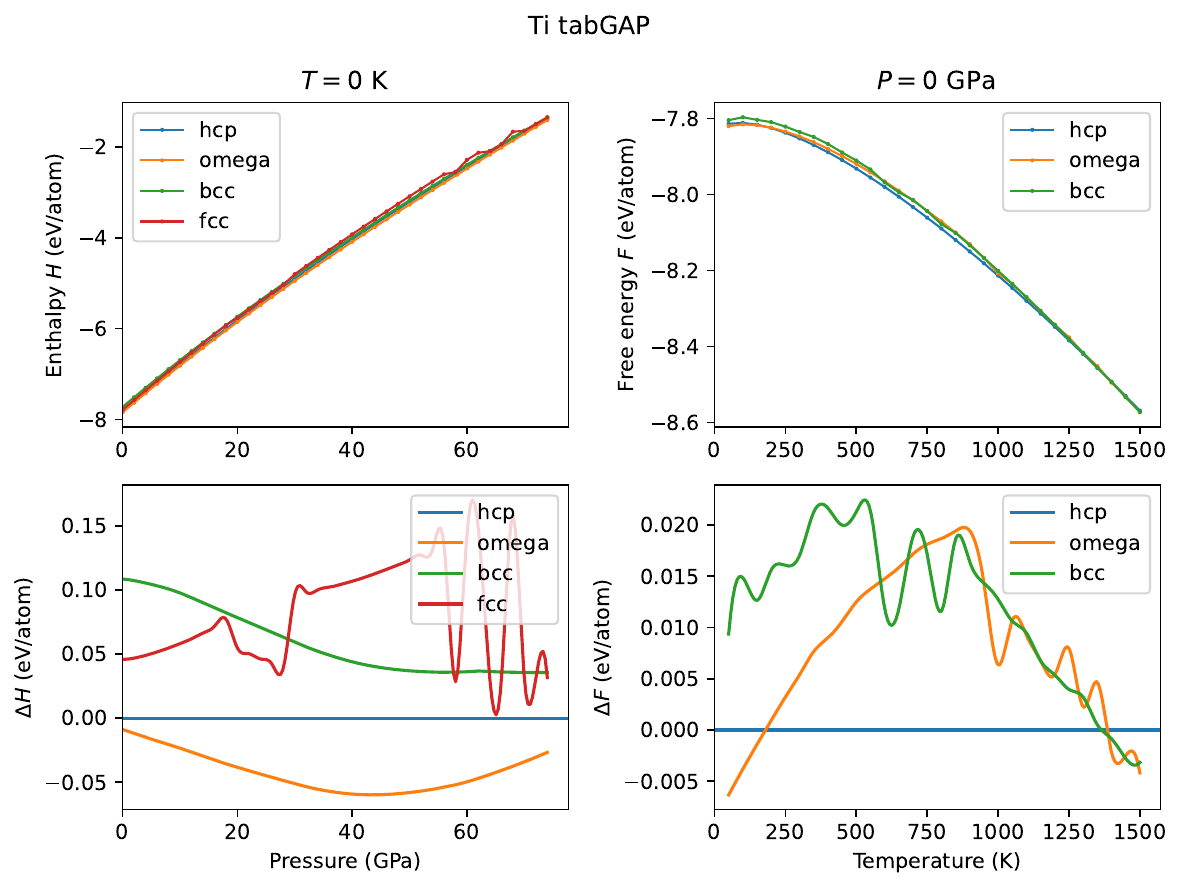}
    \includegraphics[width=0.8\linewidth]{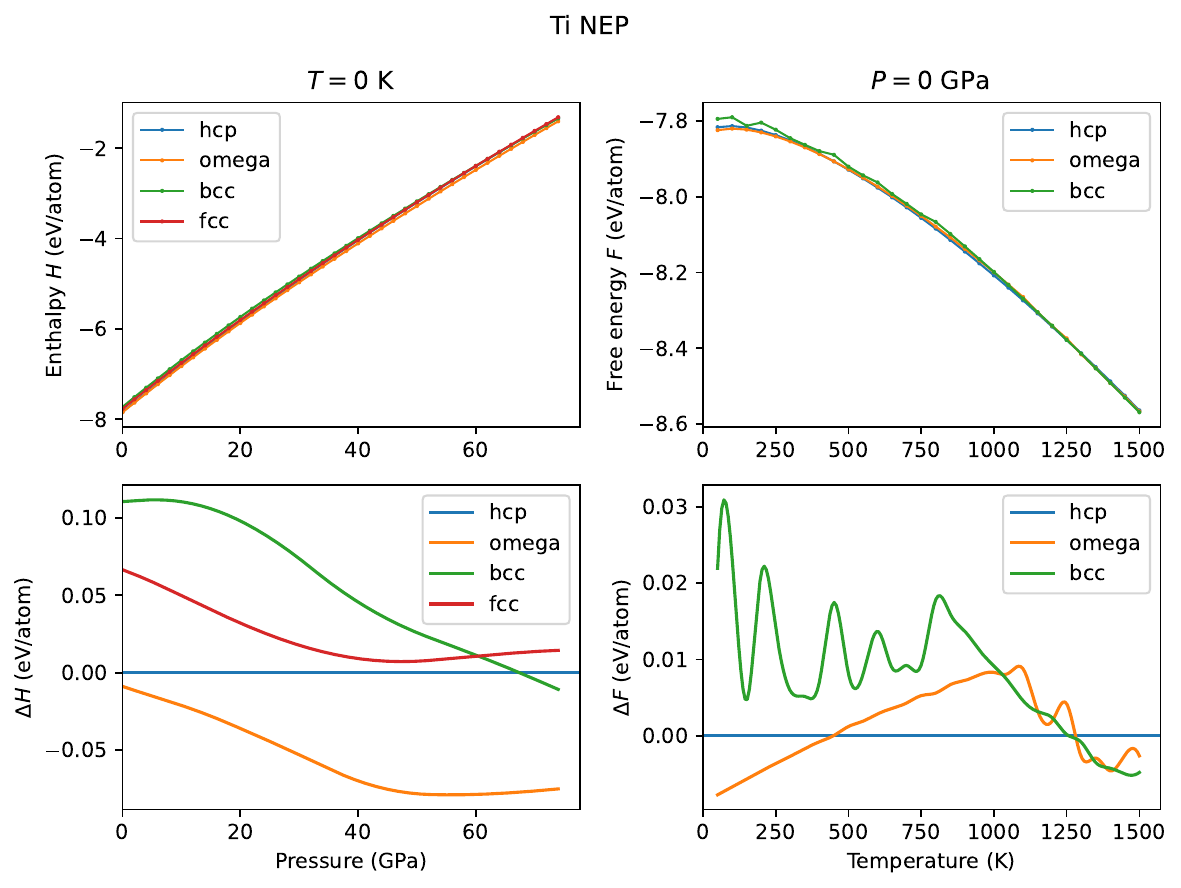}
    \caption{Simulated free energies and free-energy differences compared to the hcp phase for Ti from tabGAP and NEP. Enthalpy $H$ ($T=0$ K) and Helmholtz free energy $F$ ($P=0$ GPa).}
    \label{fig:fd-Ti}
\end{figure}

\begin{figure}
    \centering
    \includegraphics[width=0.8\linewidth]{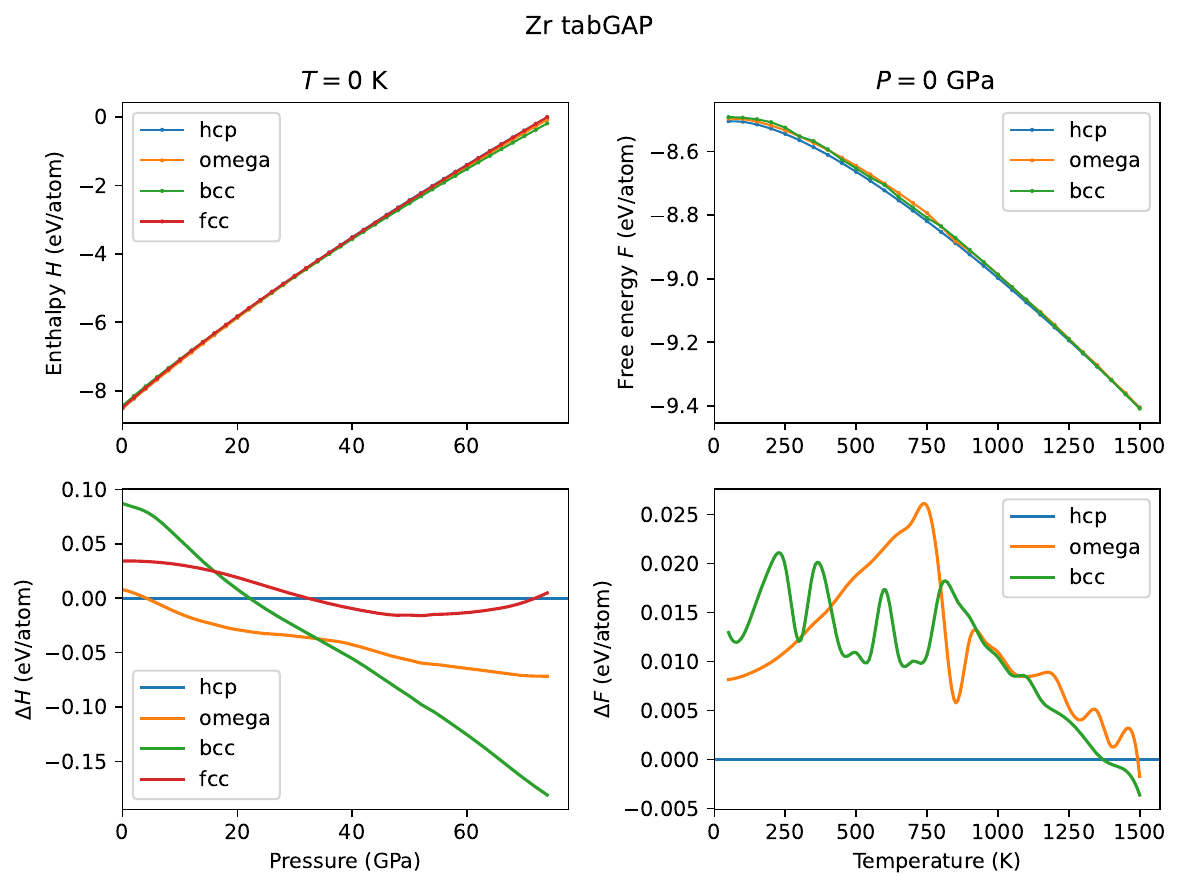}
    \includegraphics[width=0.8\linewidth]{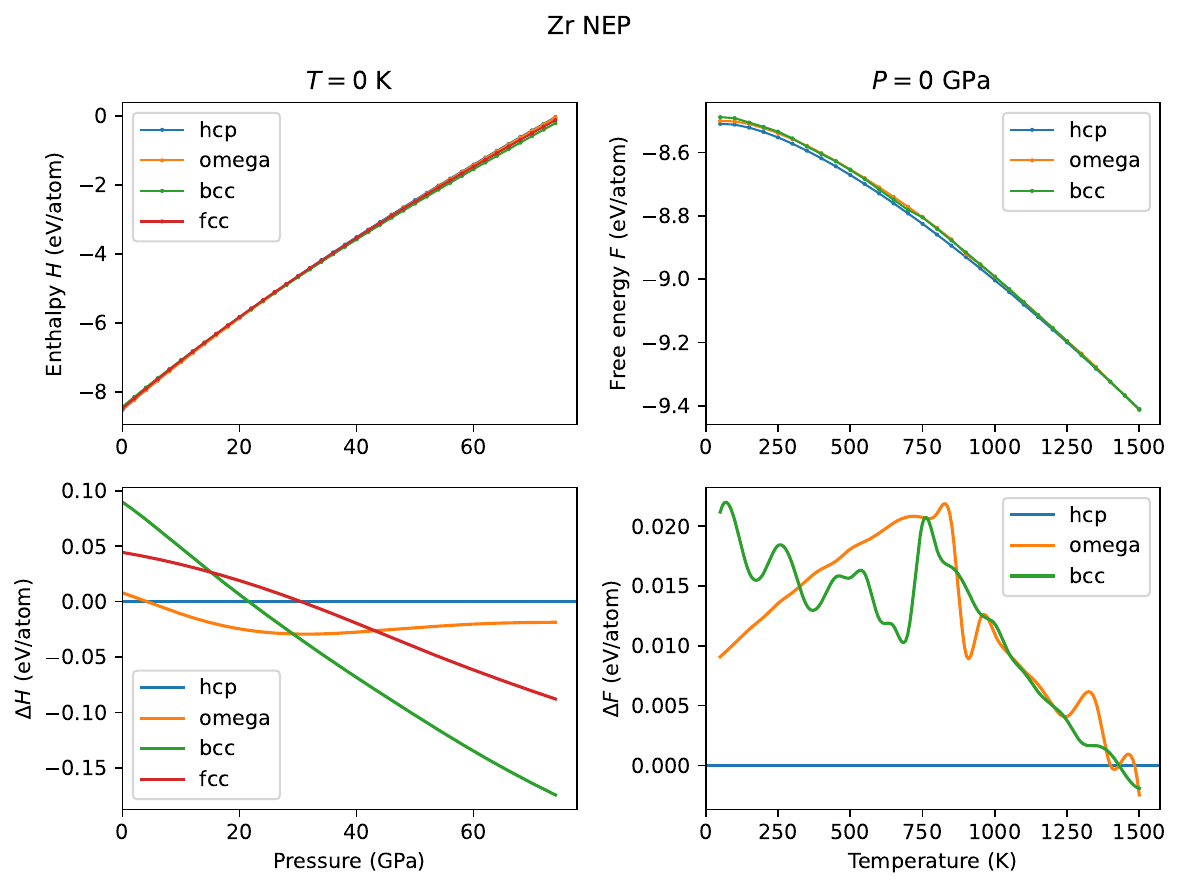}
    \caption{Simulated free energies and free-energy differences compared to the hcp phase for Zr from tabGAP and NEP. Enthalpy $H$ ($T=0$ K) and Helmholtz free energy $F$ ($P=0$ GPa).}
    \label{fig:fd-Zr}
\end{figure}

\begin{figure}
    \centering
    \includegraphics[width=0.8\linewidth]{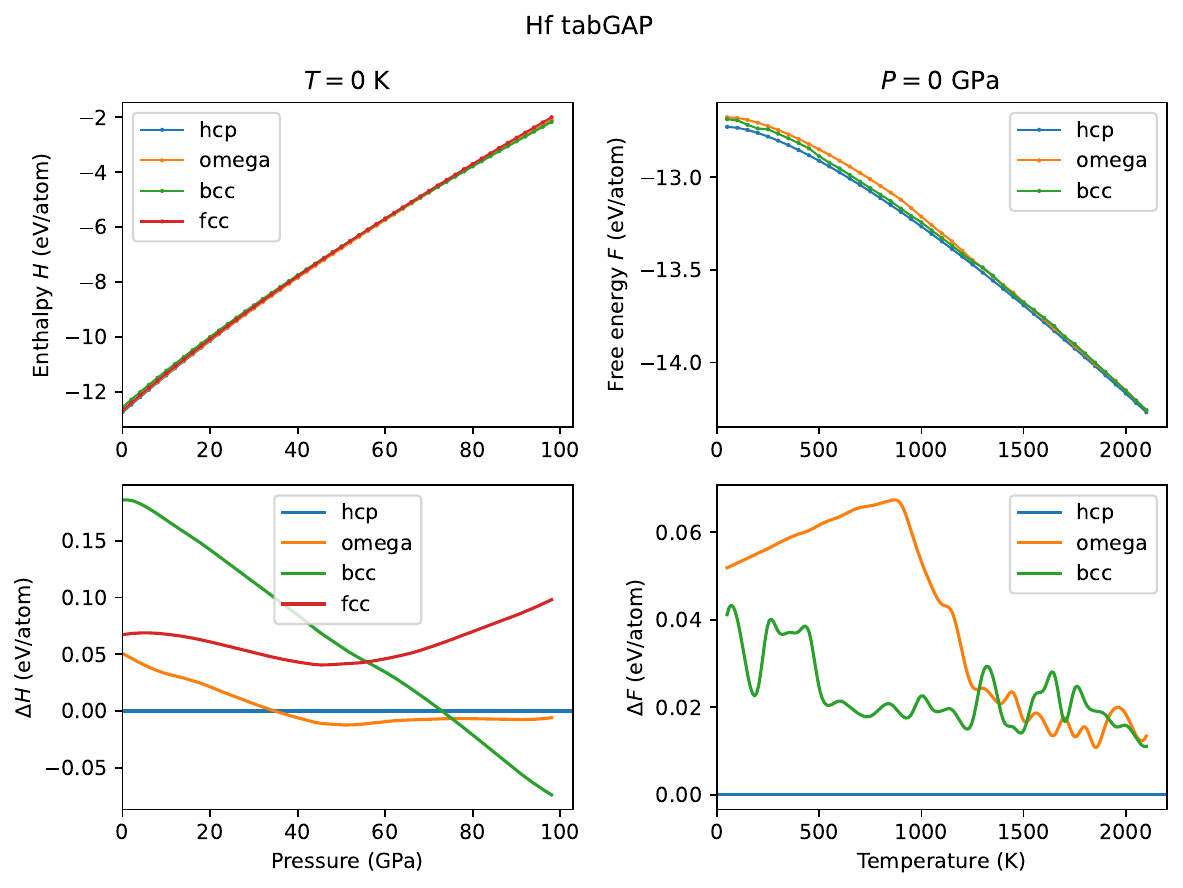}
    \includegraphics[width=0.8\linewidth]{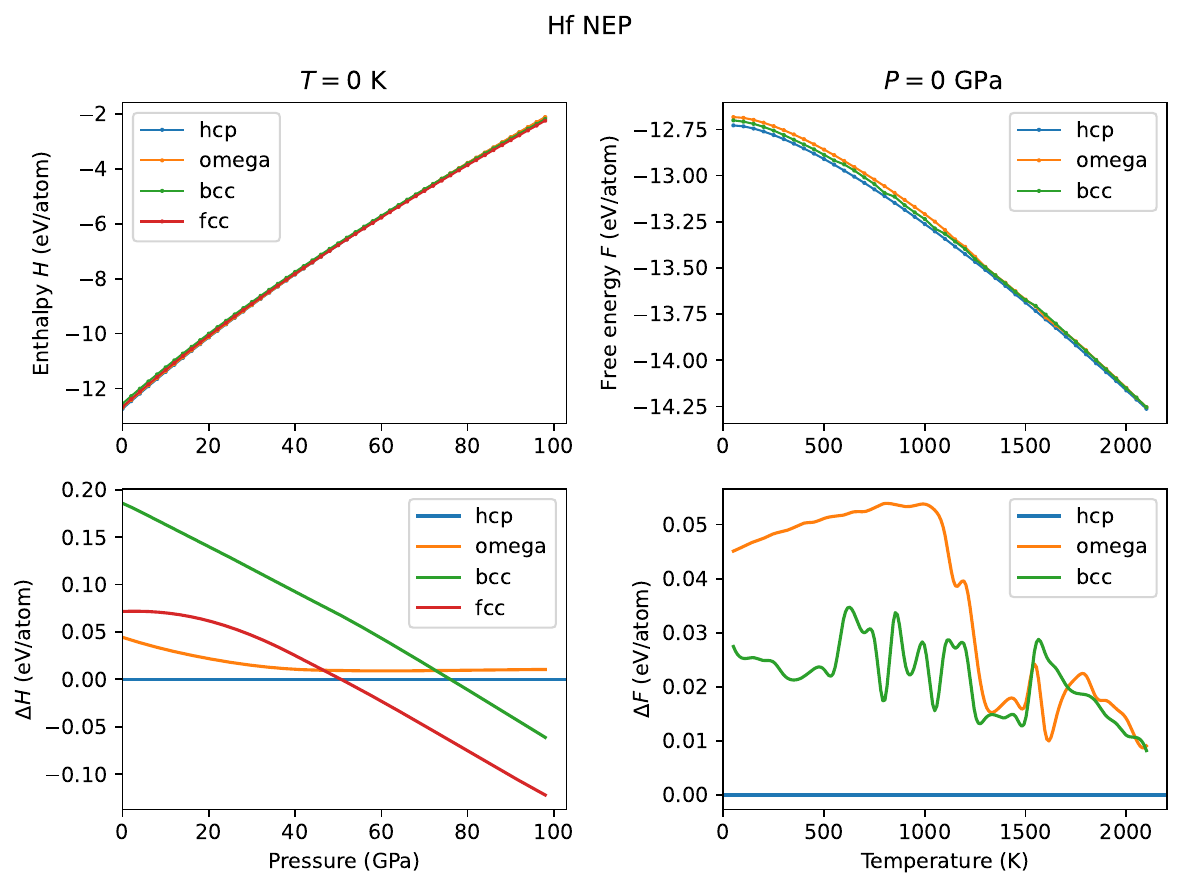}
    \caption{Simulated free energies and free-energy differences compared to the hcp phase for Hf from tabGAP and NEP. Enthalpy $H$ ($T=0$ K) and Helmholtz free energy $F$ ($P=0$ GPa).}
    \label{fig:fd-Hf}
\end{figure}

Figure~\ref{fig:SIPT} shows the phase diagrams for Ti (also in main text), Zr, and Hf simulated with tabGAP and NEP. The phase diagrams are mapped by brute-force sampling and calculation of Gibbs free energies at $(P, T)$ points using nonequilibrium MD simulations, as described in the main text. As examples, Figures~\ref{fig:fd-Ti}-\ref{fig:fd-Hf} show the simulated and calculated $G(P, T=0)$ (i.e., enthalpy $H$) and $G(T, P=0)$ (i.e., Helmholtz free energy $F$) curves for the considered phases. The same figures also show the difference in free energy compared to the hcp phase, revealing the zero-temperature and zero-pressure phase transitions shown in the phase diagrams.

The free-energy differences are computed using cubic spline interpolations of the free energy data. Note that for some phases the free energies and in particular the energy differences show artificially fluctuating curves. These fluctuations are not real and just a consequence of unstable phases partially transitioning to a more stable phase during the MD simulation. This also means they only appear for highly unstable phases and do not influence the determination of the equilibrium phase and phase transitions. The fcc phase is only included in the $T=0$ K curves in Figures~\ref{fig:fd-Ti}-\ref{fig:fd-Hf}. The fcc phase was initially not considered at all until high-pressure MD simulations of Hf with NEP showed evidence of other phases transitioning to fcc. The static (zero-temperature) enthalpy calculations revealed it to be stable at high pressures. After that, we sparsely sampled $(P, T)$ points to map the fcc phase region in Hf NEP. For all other cases, the enthalpy curves suggest that the fcc phase is always far from the most stable phase.

\clearpage